\documentclass[11pt]{article}
\pdfoutput=1

\usepackage{amssymb}
\usepackage{amsmath}
\usepackage{amsthm}
\usepackage{mathtools}
\usepackage[usenames,dvipsnames]{xcolor}
\usepackage{epsfig}
\usepackage{dcolumn}
\usepackage{tikz}
\usepackage{tikz-cd}
\usetikzlibrary{shapes.geometric, arrows}
\usepackage{upgreek}
\usepackage{setspace}
\usepackage{enumitem}
\usepackage{array,multirow,bigdelim,arydshln}
\usepackage{appendix}
\usepackage[export]{adjustbox}
\usepackage{xparse}
\usepackage[utf8]{inputenc}
\usepackage{microtype}
\usepackage{bm}
\usepackage{braket}
\usepackage{dsfont}
\usepackage{nccmath}

\usepackage{jheppub}

\usepackage{hyperref}
\hypersetup{
	colorlinks=true,
	urlcolor=Maroon,
	linkcolor=Maroon,
	citecolor=Maroon,
	bookmarks=true,
	pdfauthor={Sebastian Mizera and Andrzej Pokraka},
	pdftitle={From Infinity to Four Dimensions: Higher Residue Pairings and Feynman Integrals},
	pdfdisplaydoctitle=true,
	pdfstartview=FitH,
	backref=false,
	pagebackref=false
}

\newcommand{\WidestEntry}{$-1$}%
\newcommand{\SetToWidest}[1]{\makebox[\widthof{\WidestEntry}]{$#1$}}%

\def \be {\begin{equation}}
\def \ee {\end{equation}}
\def \nn {\nonumber}
\def \la {\langle}
\def \ra {\rangle}
\def \C {\mathbb{C}}
\def \R {\mathbb{R}}
\def \Q {\mathbb{Q}}

\def \SL {\mathrm{SL}}

\def \GL {\mathrm{GL}}

\def \CP {\mathbb{CP}}
\def \Z {\mathbb{Z}}

\def \M {\mathcal{M}}

\def \ep {\epsilon}
\def \vphi {\varphi}
\def \vep {\varepsilon}

\DeclareMathOperator{\Res}{Res}

\DeclareMathOperator{\Crit}{Crit}

\theoremstyle{definition}

\newtheorem*{outline}{Outline}

\allowdisplaybreaks
\graphicspath{{figures/}}
\renewcommand{\thefigure}{\thesection.\arabic{figure}}

\title{From Infinity to Four Dimensions:\\Higher Residue Pairings and Feynman Integrals}
\author[a]{Sebastian Mizera}\emailAdd{smizera@ias.edu}
\affiliation[a]{Institute for Advanced Study, Einstein Drive, Princeton, NJ 08540, USA}
\author[b]{and Andrzej Pokraka}\emailAdd{andrzej@physics.mcgill.ca}
\affiliation[b]{Department of Physics, McGill University, 3600 Rue University, Montr\'eal, QC, Canada H3A 2T8}

\abstract{We study a surprising phenomenon in which Feynman integrals in $D=4-2\varepsilon$ space-time dimensions as $\varepsilon \to 0$ can be fully characterized by their behavior in the opposite limit, $\varepsilon \to \infty$. More concretely, we consider vector bundles of Feynman integrals over kinematic spaces, whose connections have a polynomial dependence on $\varepsilon$ and are known to be governed by intersection numbers of twisted forms. They give rise to differential equations that can be obtained \emph{exactly} as a truncating expansion in either $\varepsilon$ or $1/\varepsilon$. We use the latter for explicit computations, which are performed by expanding intersection numbers in terms of Saito's higher residue pairings (previously used in the context of topological Landau--Ginzburg models and mirror symmetry). These pairings localize on critical points of a certain Morse function, which correspond to regions in the loop-momentum space that were previously thought to govern only the large-$D$ physics. The results of this work leverage recent understanding of an analogous situation for moduli spaces of curves, where the $\alpha' \to 0$ and $\alpha' \to \infty$ limits of intersection numbers coincide for scattering amplitudes of massless quantum field theories.}

\begin{document}

\maketitle
\setcounter{page}{2}

\section{Introduction}

One of the more surprising developments in the recent study of scattering amplitudes has been the introduction of \emph{scattering equations} \cite{Cachazo:2013gna}, which allow for writing tree-level amplitudes and loop-level integrands---such as those of Yang--Mills or gravity theories---in terms of certain localization integrals on moduli spaces of punctured Riemann spheres \cite{Cachazo:2013hca,Cachazo:2013iea}. Scattering equations can be understood as critical-point conditions for a certain ``potential'' function $W$, determined by the vanishing of its first derivative, $dW=0$.

It was later understood that such localization formulae are not at all specific to moduli spaces and can be broadly extended to more general cases \cite{Mizera:2017rqa}. To be specific, let us consider a complex manifold $M$ written as a complement of a finite number of hypersurfaces in $\CP^m$ with inhomogeneous coordinates $(z_1, z_2, \ldots, z_m)$, as well as a potential function $W(z_1,z_2,\ldots,z_m)$ with logarithmic singularities on those hypersurfaces. To two top holomorphic forms, $\varphi_- = \widehat{\varphi}_- d^m z$ and $\varphi_+ = \widehat{\varphi}_+ d^m z$, we associate a pairing, which following \cite{Mizera:2017rqa,Mizera:2019gea} we state as a Grothendieck residue around the critical points,
\be\label{K0}
\Res_{dW=0} \left( \frac{\widehat{\varphi}_- \widehat{\varphi}_+\, d^m z}{\partial_1 W\, \partial_2 W\, \cdots\, \partial_m W} \right),
\ee
where $d^m z$ is the measure form and $\partial_i = \partial/\partial z_i$. More geometrically, it should be understood as \emph{a} self-duality pairing of the cohomology of the Koszul complex $(\Omega^\bullet_M, dW\wedge)$. 
When applied to the moduli space of Riemann spheres with $n$ punctures, $M{=}\M_{0,n}$, the pairing \eqref{K0} coincides with the Cachazo--He--Yuan formula \cite{Cachazo:2013hca}.

In this work we study the connection to so-called \emph{higher residue pairings} \cite{saito1983higher}, which are a family of objects generalizing \eqref{K0}. They were introduced by Saito in the context of singularity theory, which itself aims at a higher-dimensional generalization of the classic theory of elliptic integrals \cite{saito1982periods,19831231,10004589614}. Higher residue pairings already play an important role in theoretical physics, especially in the context of mirror symmetry and topological Landau--Ginzburg models \cite{Vafa:1990mu,Cecotti:1991me,Blok:1991bi,Dijkgraaf:1991qh,dubrovin1998painleve,Chiodo_2013,Li:2013kja,Saito:2014oda,Li:2014eua,Lerche:2018gvn,Li:2019qrl}, conformal field and string theories \cite{Blok:1991bi,Dubrovin:1994hc,Losev:1992tt,Losev:1998dv,Belavin:2015wxa}, and Seiberg--Witten theory \cite{Li:2018rdd}, among others. Based on this list a connection to scattering amplitudes of ``garden-variety'' quantum field theories might already sound rather surprising.

The first correction to \eqref{K0} is given by the higher residue pairing \cite{saito1983higher}
\be\label{K1}
\Res_{dW=0} \left( \frac{1}{2} \sum_{i=1}^{m} \frac{( \widehat{\varphi}_+\, \partial_i \widehat{\varphi}_- - \widehat{\varphi}_-\, \partial_i \widehat{\varphi}_+) \,d^m z}{\partial_1 W \cdots (\partial_i W)^2 \cdots \partial_m W} \right).
\ee
Notice that it has $m{+}1$ powers of $W$ in the denominator, compared to just $m$ in \eqref{K0}. This motivates an introduction of a book-keeping parameter $\tau$ and sending $W \to \tau W$ (in quantum field theory $\tau$ is proportional to the inverse of Planck's constant $\hbar^{-1}$, in string theory it is the inverse string tension $\alpha'$, while for Feynman multi-loop integrals it becomes the dimension-regularization parameter $\varepsilon$). All higher residue pairings may be compactly written as a $\tau^{-1}$ expansion of a single object,
\be\label{Kall}
\la\varphi_- | \varphi_+\ra_{dW} = \eqref{K0} + \tau^{-1} \eqref{K1} + \cdots,\nn
\ee
which in fact gives a compact expression that generates all-order corrections. Geometrically \eqref{Kall} is the \emph{intersection number} of cohomology classes associated to the twisted de Rham complexes $(\Omega^\bullet_M, d{\pm} \tau dW\wedge)$, see, e.g., \cite{zbMATH03996010,cho1995}, which will be reformulated in terms of a {\v C}ech--de Rham double complex later in the text.

The physical meaning of intersection numbers on $M{=}{\cal M}_{0,n}$ (with $\tau = \alpha'$) is that they compute tree-level scattering amplitudes of quantum field theories with a finite spectrum of masses, $m^2 \in \Z/\alpha'$ \cite{Mizera:2017rqa,Mizera:2019gea}, which are rational functions of kinematic invariants with simple poles of the form $\frac{1}{p^2 + \Z/\alpha'}$. As a matter of fact, they were used to resolve a long-standing puzzle regarding scattering equations, which---despite computing \emph{low-energy} physics \cite{Cachazo:2013hca}---determine worldsheets dominating in the \emph{high-energy} limit of string theory \cite{Gross:1987kza,Gross:1987ar}.\footnote{One of the main sources of this confusion was the fact that the $\alpha' \to \infty$ limit of string amplitudes was often stated incorrectly in the literature as being dominated by a finite number of saddle points. For this reason we review it in Appendix~\ref{app:Gross-Mende} in the simplest case of $n{=}$4 at tree-level.} On the one hand, in the $\alpha' \to 0$ limit intersection numbers coincide with the low-energy limit of string-theory scattering amplitudes. On the other hand, in the $\alpha' \to \infty$ limit they reproduce the localization on scattering equations. One may ask when the two limits agree. This clearly happens when the intersection number is independent (or homogeneous) of $\alpha'$ in the first place \cite{Mizera:2017rqa,Mizera:2019gea}, as then it does not matter if we send $\alpha' \to 0$ or $\alpha' \to \infty$! Physically, this property corresponds to propagators of the form $\frac{1}{p^2}$, i.e., when intersection numbers compute amplitudes of \emph{massless} quantum field theories. 

The main goal of this paper is to leverage this new understanding to other problems in scattering amplitudes.

In particular, we focus on multi-loop Feynman integrals in $D=4{-}2\varepsilon$ space-time dimensions, as those have a known interpretation in the same geometric language \cite{Mastrolia:2018uzb}.\footnote{Another direction in which one can generalize this formalism is from ${\cal M}_{0,n} = \mathrm{G}(2,n)/\mathrm{T}^{n-1}$ to more general maximal torus quotients of Grassmannians, $\mathrm{G}(k,n)/\mathrm{T}^{n-1}$, as initiated recently in \cite{Cachazo:2019ngv}.} The reason for employing dimensional regularization is that such integrals most often do not converge in strictly four dimensions. They can be written as
\be\label{intro-I}
I = \int_{\Gamma} e^{\varepsilon W}\, {\varphi}_+,
\ee
where $\Gamma$ is some integration cycle and the potential $W$ is determined in terms of so-called Symanzik polynomials that specify the topology of a given graph $G$. In this case the role of $\tau$ is played by $\varepsilon$ and $M{=}{\cal M}_G$ is the moduli space of Riemannian metrics on $G$ with coordinates given by Schwinger parameters.

Alternatively, Feynman integrals can be understood as sections of vector bundles over the kinematic space, defined by the solutions of the system of differential equations
\be\label{intro-DE}
({\cal D} - {\mathbf\Omega}\wedge ) \vec{I} = 0,
\ee
where ${\vec I}$ is a vector of integrals of the type \eqref{intro-I}, ${\cal D}$ is the differential on the kinematic space, and $\mathbf{\Omega}$ is a matrix-valued one-form, subject to integrability constraints, that needs to be determined. Together with boundary conditions, which we assume are known, \eqref{intro-DE} fully characterizes the behavior of Feynman integrals in a given family around $\varepsilon \to 0$. Thus the problem amounts to finding the matrix $\mathbf{\Omega}$. It was recently realised that fibers of the vector bundle can be described by the cohomology of $(\Omega^\bullet_M, d{+} \tau dW\wedge)$ and hence the entries of $\mathbf{\Omega}$ can be computed by the same intersection numbers \eqref{Kall} described above \cite{Mastrolia:2018uzb,Frellesvig:2019uqt}.

As a matter of fact, on physical grounds $\mathbf{\Omega}$ must be a \emph{polynomial} in $\varepsilon$, as it can be shown that any pole in $\varepsilon$ must be spurious (see, e.g., \cite{Henn:2014qga}),
\be
\mathbf{\Omega} = \sum_{k=0}^{k_{\text{max}}} \varepsilon^k\, \mathbf{\Omega}_{(k)}.
\ee
Because of this we can expand the matrix $\mathbf{\Omega}$ around either $\varepsilon\to 0$ or $\varepsilon \to \infty$ and still obtain the \emph{exact} result with a finite number of terms! We use the latter option, which combined with the expansion \eqref{Kall} allows us to compute $\mathbf{\Omega}$ in terms of higher residue pairings $\eqref{K0}$, $\eqref{K1}$, and their further corrections. Notice that critical points contributing to these computations correspond to places on the moduli space ${\cal M}_G$ that normally dominate the $\varepsilon \to \infty$ physics. This is yet another example of what seems to be a more general moduli space localization phenomenon \cite{Mizera:2019gea}, in which physical quantities in one limit can be extracted from the exact opposite one.

We illustrate this new idea by performing explicit computations for two families of integrals. We start with arguably the simplest case of a single-box massless integral and follow with a two-loop sunrise diagram with masses running in the loops.

\begin{outline}
In Section~\ref{sec:geometric-setup} we review the geometric setup based on twisted de Rham and {\v C}ech--de Rham cohomologies, which leads to explicit expressions for higher residue pairings. In Section~\ref{sec:from-infinite-to-four-dimensions} we formulate Feynman integrals as twisted periods and describe how to obtain their differential equations from higher residue pairings. Explicit examples are given in Section~\ref{sec:massless-box} for the one-loop box diagram and in Section~\ref{sec:massive-sunrise} for the two-loop massive sunrise diagram. We conclude in Section~\ref{sec:discussion} with a discussion of future directions. This paper comes with Appendix~\ref{app:Gross-Mende}, where we clarify the computation of $\alpha' \to \infty$ asymptotics of genus-zero string amplitudes.
\end{outline}

\section{\label{sec:geometric-setup}Geometric Setup}

In this section we briefly review the geometric setup underlying the remainder of the paper. The understanding of Sections~\ref{sec:twisted-complex}--\ref{sec:Cech-de-Rham} is not needed to compute higher residue pairings in practice, but rather is meant to give an intuition about where they come from. Explicit expressions for higher residue pairings are given in Section~\ref{sec:higher-residue-pairings}, and the way of relating them to integrals over middle-dimensional cycles is explained in Section~\ref{sec:integration}.

\subsection{\label{sec:twisted-complex}From Koszul to Twisted de Rham Complex}

The formula \eqref{K0} can be understood geometrically in the following way \cite{Mizera:2019gea}. Let us consider $M = \CP^m - \cup_{i=1}^{k} H_i$, where each $H_i$ is a hypersurface in $\CP^m$. Integrals defined on such spaces are ubiquitous in physics, e.g., in Feynman multi-loop integrals or string perturbation theory.

We introduce a holomorphic function $W$ on the covering space $\widehat{M}$ of $M$ with logarithmic singularities along each $H_i$. For instance, if $H_i$ are defined by equations $\{f_i = 0\}$ then
\be
W = \sum_{i=1}^{k} \alpha_i \log f_i
\ee
for generic constants $\alpha_i$, with $\sum_{i=1}^{k} \alpha_i = 0$, is a valid choice of $W$.\footnote{Alternatively we could have worked on $M = \C^m - \cup_{i=1}^{k} H_i$ with the constraint $\sum_{i=1}^{k} \alpha_i = 0$ lifted.} We will often call $W$ a \emph{potential} to use the same nomenclature as in the literature on mirror symmetry \cite{hori2003mirror}. Let us consider a single-valued holomorphic one-form $dW$. Defining $\Omega^k_M$ to be the space of smooth $k$-forms on $M$ (with ${\cal O}_M := \Omega^0_M$ being the space of functions), we introduce the following sequence:
\be\label{Koszul}
\begin{tikzcd}
0 \arrow[r] & {\mathcal O}_M \arrow[r, "dW\wedge"] & \Omega_M^1 \arrow[r, "dW\wedge"] & \cdots \arrow[r, "dW\wedge"] & \Omega_M^m \arrow[r] & 0,
\end{tikzcd}
\ee
called the \emph{Koszul cochain complex} $(\Omega^\bullet_M, dW\wedge)$. Here each map is given simply by wedging the element $\varphi_k \in \Omega^k_M$ from the left with $dW\wedge$, that is
\be
dW\wedge:\; \varphi_k \;\mapsto\; dW\wedge \varphi_k
\ee
for $k=0,1,\ldots,m{-}1$. Since $dW\wedge dW =0$, the sequence \eqref{Koszul} is exact, meaning that image of each map is equal to the kernel of the following one (applying two consecutive maps gives zero). This allows us to construct cohomology groups $H^k(M,dW\wedge)$ associated to \eqref{Koszul}, which are given by kernel of each $dW\wedge$ modulo the image of the preceding $dW\wedge$, or in other words
\be\label{dW-cohomology}
H^{k}(M,dW\wedge) := \frac{\{ \varphi_k \in \Omega^k_M \,|\, dW\wedge \varphi_k = 0 \}}{\{dW\wedge \varphi_{k-1} \in \Omega^k_M \,|\, \varphi_{k-1} \in \Omega^{k-1}_M \}}.
\ee
One can show that only the case $k=m$ is non-trivial, provided the constants $\alpha_i$ are generic \cite{Esnault1992}. In addition, since $dW\wedge$ defines a rank-$1$ flat connection we have
\be\label{dim-Hm}
\dim H^m(M,dW\wedge) = (-1)^{m} \chi(M),
\ee
which allows us to compute the dimension of the above cohomology group purely topologically in terms of the Euler characteristic $\chi(M)$ of $M$.

From now on we assume that $\Re(W)$ is a \emph{Morse function} \cite{milnor2016morse} with isolated and non-degenerate critical points. It is easily seen that such critical points are given by $dW=0$, i.e., coincide with the critical points of the potential function $W$. For later convenience let us introduce notation for the critical locus of $W$:
\be
\Crit(W) := \{ (z_1, z_2, \ldots, z_m) \in M \,|\, dW= 0 \},
\ee
which by the above assumptions is a finite set. Since $W$ is holomorphic, all critical points have the same Morse index, i.e., the same number of independent upwards and downwards directions extending from it. This tells us that \cite{aomoto1987gauss,Silvotti1996}
\be
\#\Crit(h) = (-1)^m \chi(M),
\ee
which combined with \eqref{dim-Hm} allows one to compute the dimension of $H^m(M,dW\wedge)$ by counting critical points.

One can define a self-duality pairing of $H^m(M,dW\wedge)$, which is given by \cite{Mizera:2017rqa}
\be\label{residue-pairing}
(\varphi_- | \varphi_+)_{dW,0} \,:=\, \Res_{dW=0} \left( \frac{\widehat{\varphi}_- \widehat{\varphi}_+ d^m z}{ \partial_1 W \, \partial_2 W\, \cdots\, \partial_m W} \right)
\ee
for $\varphi_\pm \in H^m(M,dW\wedge)$. Here the hat denotes stripping an overall differential from a form, $\widehat{\varphi}\, d^m z := \varphi$ and $d^m z = \wedge_{i=1}^{m} dz_i$. Alternatively we can think of the hatted function as being defined in the ring of functions modulo the ideal generated by $\partial_i W = 0$,
\be
\widehat{\varphi}_\pm \in {\cal O}_{M}/(\partial_1 W, \partial_2 W, \ldots, \partial_m W).
\ee
The symbol $\Res_{dW=0}$ denotes a sum over Grothendieck residues around each critical point \cite{hartshorne2014residues}, which is simply given by
\be\label{residue-def}
\Res_{dW=0}\left( \eta \right) := \frac{1}{(2\pi \sqrt{-1})^m} \oint_{|\partial_1 W| = \varepsilon} \oint_{|\partial_2 W|  = \varepsilon} \!\!\!\cdots\; \oint_{|\partial_m W| = \varepsilon} \eta.
\ee
Here the contour is oriented by $d(\arg \partial_1 W)\wedge \cdots \wedge d(\arg \partial_m W) > 0$ and has support only on small tubular neighbourhoods are each critical point. It will be evaluated directly in many situations later in the text.

It is important to note that the pairing \eqref{residue-pairing} is not unique. As a matter of fact, we can embed it into the following formalism. Consider the exact sequence
\be\label{twisted-complex}
\begin{tikzcd}
0 \arrow[r] & {\mathcal O}_M \arrow[r, "\nabla_{dW}"] & \Omega_M^1 \arrow[r, "\nabla_{dW}"] & \cdots \arrow[r, "\nabla_{dW}"] & \Omega_M^m \arrow[r] & 0
\end{tikzcd}
\ee
called a \emph{twisted de Rham complex}, where we introduced a differential
\be
\nabla_{dW} := d + \tau dW\wedge,
\ee
which defines an integrable connection, or equivalently a flat line bundle, since $\nabla_{dW}^2 =0$. As was the case before, we can define cohomology groups based on this complex\footnote{Since $M$ is finite-dimensional, one can equivalently think of our setup in terms of Batalin--Vilkovisky (BV) formalism, see, e.g., \cite{Witten:1990wb}. More precisely, for a space $\text{MV}_\bullet(M) := \Gamma(\wedge^\bullet TM)$ of antisymmetric multivector fields on $M$ (a counterpart of $\Omega^\bullet(M) = \Gamma(\wedge^\bullet T^\ast M)$) and a BV differential $\partial_{dW} := \text{div} + \iota_{dW}$, the corresponding BV complex $(\text{MV}_\bullet(M), \partial_{dW})$ is isomorphic to $(\Omega^{m-\bullet}(M),\nabla_{dW})$.}
\be\label{twisted-cohomology}
H^{k}(M,\nabla_{dW}) := \frac{\{ \varphi_k \in \Omega^k_M \,|\, \nabla_{dW} \varphi_k = 0 \}}{\{\nabla_{dW} \varphi_{k-1} \in \Omega^k_M \,|\, \varphi_{k-1} \in \Omega^{k-1}_M \}},
\ee
which are spaces of $\nabla_{dW}$-closed modulo $\nabla_{dW}$-exact forms. We will call $H^k_{dW} := H^{k}(M,\nabla_{dW})$ for short from now on. As in the case of \eqref{dW-cohomology}, only $k=m$ gives a non-trivial cohomology \cite{aomoto1975vanishing} and hence we have
\be
\dim H^m_{dW} = (-1)^m \chi(M).
\ee
Let us introduce a dual twisted cohomology $H^m_{-dW}$ defined in the same way as $H^m_{dW}$ but with $W \to -W$. We will often refer to cohomology classes $[\varphi_\pm] \in H^{m}_{\pm dW}$ as \emph{twisted cocycles} and specific representatives $\varphi_\pm$ as \emph{twisted forms}. Duality of the two cohomologies is induced by the intersection pairing
\be
H^m_{-dW} \times H^m_{dW} \to \C
\ee
defined by
\be\label{intersection-definition}
\la \varphi_- | \varphi_+ \ra_{dW} \,:=\, \left(\frac{\tau}{2\pi \sqrt{-1}}\right)^m \int_{M} \varphi_- \wedge \varphi_+^c
\ee
and called an \emph{intersection number}. The overall normalization is chosen for later convenience. Since $M$ is non-compact, for this definition to make sense one needs to use a compactly-supported form $\varphi_+^c$ (i.e., one which vanishes in infinitesimal neighbourhoods of the boundary divisor $\partial M$) in the same cohomology class as $\varphi_+$ for direct computations. This not only makes the result depend on the potential $W$, but also means the above integral localizes on $\partial M$, as in the bulk of $M$ we have $\varphi_- \wedge \varphi_+^c = \varphi_- \wedge \varphi_+ = 0$ for two top holomorphic forms, see, e.g., \cite{Mizera:2019gea}.

Intersection numbers are rational functions of $\alpha_i$'s and $\tau$. Different ways of evaluating them in practice were given in \cite{cho1995,matsumoto1994,matsumoto1998,Matsumoto1998-2,majima2000,OST2003,doi:10.1142/S0129167X13500948,goto2015,goto2015b,Yoshiaki-GOTO2015203,Mizera:2017rqa,Mizera:2019gea}. For spaces $M$ admitting a fiber bundle decomposition (or, more precisely, such that the connection decomposes generically on the fibers), the most efficient computation method is currently given by recursion relations \cite{Mizera:2019gea}. When $M={\cal M}_{0,n}$ is the moduli space of genus-zero curves with $n$ marked points, intersection numbers have an intrinsic interpretation as computing tree-level scattering amplitudes of quantum field theories \cite{Mizera:2017rqa,Mizera:2019gea}. Worldsheet models that might underlie these computations were studied in \cite{Siegel:2015axg,Lee:2017utr,Casali:2017mss,Jusinskas:2019dpc}. In those cases the corresponding potential $W$ can be given an interpretation as an electrostatic potential for a system of particles \cite{Cachazo:2016ror}.

In the context of Landau--Ginzburg models, intersection numbers between basis elements compute entries of a metric $g_{ij} = \la \varphi_i | \varphi_j \ra_{dW}$ on the space of physical operators, see, e.g., \cite{Blok:1991bi}. A particular problem, related to the theory of Frobenius manifolds, is finding bases that make this metric flat, i.e., $g_{ij} = \delta_{ij}$.

Since \eqref{dW-cohomology} looks like a limit $\tau \to \infty$ of \eqref{twisted-cohomology} it is natural to expect that the residue pairing $(\varphi_- | \varphi_+)_{dW,0}$ should be related to the limit of the intersection number $\la \varphi_- | \varphi_+ \ra_{dW}$. Before deriving this result in full generality, let us consider the one-dimensional case $\dim_{\C} M = 1$ to gain some intuition about this relationship.

\subsection{\label{sec:one-dimensional}Example: One-Dimensional Case}

Let us consider a one-dimensional case, where each hypersurface $H_i$ is a single point, say $\{z=p_i\}$, removed from $\CP^1$,
\be
M = \CP^1 - \{p_1,p_2,\ldots,p_k\}
\ee
and the corresponding potential is $W = \sum_{i=1}^{k} \alpha_i \log(z{-}p_i)$ with $\alpha_i$'s adding up to zero. The Euler characteristic is simply $\chi(M) = 2-k$, which is the same as the number of critical points, as the solutions of $dW=0$ are roots of a degree-$(k{-}2)$ polynomial in $z$.

The intersection number of two twisted forms $\varphi_- \in H^1_{-dW}$ and $\varphi_+  \in H^1_{dW}$ is defined by
\be\label{intersection-number-1d}
\la \varphi_- | \varphi_+ \ra_{dW} = \frac{\tau}{2\pi \sqrt{-1}} \int_M \varphi_- \wedge \varphi_+^c.
\ee
Let us construct $\varphi_+^c$ explicitly as
\be
\varphi_+^c = \varphi_+ - \nabla_{dW} \left( \sum_{i=1}^k \Theta(|z{-}p_i|^2 {-} \varepsilon^2)\, \nabla_{dW}^{-1} \varphi_+ \right),
\ee
which is manifestly cohomologous to $\varphi_+$. Here $\Theta(x)$ is a step function equal to one for $x{>}0$ and zero otherwise. In this way, each term in the sum has support on an infinitesimal disk around $p_i$ with radius $\varepsilon$. The inverse differential $\nabla_{dW}^{-1}$ is defined such that $\nabla_{dW} \nabla_{dW}^{-1} \eta = \eta$. In particular, $\nabla_{dW}^{-1} \varphi_+$ is a zero-form. Let us check that the resulting form has compact support by expanding the above expression:
\be\label{varphi-c}
\varphi_+^c = \varphi_+ \left( 1- \sum_{i=1}^k \Theta(|z{-}p_i|^2 {-} \varepsilon^2) \right) - \sum_{i=1}^k \delta(|z{-}p_i|^2 {-} \varepsilon^2)\, \nabla_{dW}^{-1} \varphi_+.
\ee
The first term vanishes inside small disks around each $p_i$ and the second term has only support on the circles with radii $\varepsilon$ imposed by Dirac delta functions $\delta(x)$. Therefore $\varphi_+^c$ has compact support. One could have performed the same computation in a smooth way with bump functions instead of step functions, leading to the same final result, see, e.g., \cite{matsumoto1994}, but will not do it here for the sake of clarity.

Plugging \eqref{varphi-c} back into \eqref{intersection-number-1d}, the first term wedges to zero and only the second contribution survives, which straightforwardly expresses the intersection number as a sum of $k$ residues around each point removed from $M$:
\be\label{intersection-number-1d-result}
\la \varphi_- | \varphi_+ \ra_{dW} = -\tau \sum_{i=1}^{k} \Res_{z=p_i} \left( \varphi_- \nabla_{dW}^{-1} \varphi_+ \right).
\ee
Note that because of the residue, $\nabla_{dW}^{-1} \varphi_+$ needs to be computed only locally as a holomorphic expansion around the each $p_i$ to some finite order in $z{-}p_i$ (depending on the order of the pole of $\varphi_-$). Simple power counting reveals that a given boundary at $\{z=p_i\}$ gives non-zero contribution only when the orders of poles of $\varphi_-$ and $\varphi_+$ at this point add up to at least two. Of course, we could have imposed compact support on $\varphi_-$ instead, which after an analogous computation gives a different representation
\be
\la \varphi_- | \varphi_+ \ra_{dW} = \tau \sum_{i=1}^{k} \Res_{z=p_i} \left( \varphi_+ \nabla_{-dW}^{-1} \varphi_- \right).
\ee
It will later turn out to be convenient to symmetrize between the two type of expressions, but for the time being let us stick with \eqref{intersection-number-1d-result}.

In order to make connections to the residue pairings, let us expand the inverse of the twisted differential $\nabla_{dW}^{-1}$ in powers of $\tau^{-1}$,
\be\label{inverse-nabla}
\nabla^{-1}_{dW} \varphi_+ = \tau^{-1} \frac{\widehat\varphi_+}{\partial_z W} - \tau^{-2} \frac{1}{\partial_z W} \partial_z\! \left( \frac{\widehat\varphi_+}{\partial_z W} \right) + \tau^{-3} \frac{1}{\partial_z W} \partial_z\! \left( \frac{1}{\partial_z W} \partial_z\! \left( \frac{\widehat\varphi_+}{\partial_z W} \right) \right) - \dots,
\ee
where $\partial_z := \partial/\partial z$. This expression can be confirmed by imposing $\nabla_{dW} \nabla_{dW}^{-1} \varphi_+ = \varphi_+$ order-by-order in $\tau^{-1}$. Substituting this expansion into \eqref{intersection-number-1d-result}, we can notice two facts. The first one is that $k{-}2$ new poles, at the positions of each critical point, have been introduced to the argument of the residue. Secondly, argument of each residue is now the same one-form. This allows us to deform the original contour from enclosing $\partial M = \cup_{i=1}^{k} \{z = p_i\}$ to enclosing the set of critical points $\Crit(W)$ by the residue theorem. Therefore we obtain
\be\label{residue-formula-m1}
\la \varphi_- | \varphi_+ \ra_{dW} = \tau \Res_{dW = 0} \left( \varphi_- \nabla_{dW}^{-1} \varphi_+ \right),
\ee
where $\nabla_{dW}^{-1}$ is understood as an expansion in \eqref{inverse-nabla}. We can now start collecting terms proportional to different powers of $\tau^{-1}$,
\be\label{tau-expansion}
\la \varphi_- | \varphi_+ \ra_{dW} =: \sum_{k=0}^{\infty} \tau^{-k}\, (\varphi_- | \varphi_+ )_{dW,k},
\ee
where we assumed for simplicity that $\varphi_\pm$ themselves are independent of $\tau$.
For example, the leading term is given by
\be
( \varphi_- | \varphi_+ )_{dW,0} = \Res_{dW = 0} \left( \frac{\widehat{\varphi}_- \widehat{\varphi}_+\, dz}{\partial_z W} \right),
\ee
which coincides with \eqref{residue-pairing} for $m=1$. As a matter of fact, we have an infinite number of corrections $(\varphi_- | \varphi_+)_{dW,k}$ given in \eqref{tau-expansion}, which can be straightforwardly obtained by expanding \eqref{residue-formula-m1} to higher orders. These are the simplest examples of higher residue pairings \cite{saito1983higher}.

The above example motivates looking for generalizations to higher-degree forms. In principle one should be able to carry out a similar derivation, starting with the integral expression \eqref{intersection-definition} and showing that it localizes on $\Crit(W)$ using a global residue theorem, though this path requires an involved computation. Fortunately, we can circumvent it by a change of perspective, by considering an extension of the twisted de Rham complex.

\subsection{\label{sec:Cech-de-Rham}Twisted {\v C}ech--de Rham Complex}

In this section we give an alternative, though equivalent, definition of intersection numbers that evaluates directly to the localization formula \eqref{residue-pairing} and all its $\tau^{-1}$ corrections. We follow the work of Saito \cite{saito1982periods,saito1983higher,10004589614,19831231} (for reviews see, e.g., \cite{namikawa1983,Matsuo1998}) in a language adapted to the present context.

We start by introducing a (locally-finite) open cover $\mathfrak{U} = \{U_{i+1}\}_{i=0}^{m-1}$ of the manifold $M{-}\Crit(W)$ with
\be
U_i := M - \{\partial_i W = 0\}.
\ee
It allows us to define a double complex $(C^\bullet(\mathfrak{U},\Omega^\bullet_M),\delta,\nabla_{dW})$ called the \emph{twisted {\v C}ech--de Rham complex}, which is an extension of \eqref{twisted-complex},
\be\label{Cech-de-Rham}
\begin{tikzcd}
            & 0                                                                               & 0                                                                               &                            & 0                                                                         &   \\
0 \arrow[r] & {C^0(\mathfrak{U},\Omega_M^m)} \arrow[r, "\delta"] \arrow[u]                    & {C^1(\mathfrak{U},\Omega_M^m)} \arrow[u] \arrow[r, "\delta"]                    & \cdots \arrow[r, "\delta"] & {C^{m-1}(\mathfrak{U},\Omega_M^m)} \arrow[u] \arrow[r]                    & 0 \\
            & \vdots \arrow[u, "\nabla_{dW}"']                                                & \vdots \arrow[u, "\nabla_{dW}"']                                                &                            & \vdots \arrow[u, "\nabla_{dW}"']                                          &   \\
0 \arrow[r] & {C^0(\mathfrak{U},\Omega_M^1)} \arrow[r, "\delta"] \arrow[u, "\nabla_{dW}"']    & {C^1(\mathfrak{U},\Omega_M^1)} \arrow[u, "\nabla_{dW}"'] \arrow[r, "\delta"]    & \cdots \arrow[r, "\delta"] & {C^{m-1}(\mathfrak{U},\Omega_M^1)} \arrow[r] \arrow[u, "\nabla_{dW}"']    & 0 \\
0 \arrow[r] & {C^0(\mathfrak{U},\mathcal{O}_M)} \arrow[r, "\delta"] \arrow[u, "\nabla_{dW}"'] & {C^1(\mathfrak{U},\mathcal{O}_M)} \arrow[u, "\nabla_{dW}"'] \arrow[r, "\delta"] & \cdots \arrow[r, "\delta"] & {C^{m-1}(\mathfrak{U},\mathcal{O}_M)} \arrow[r] \arrow[u, "\nabla_{dW}"'] & 0 \\
            & 0 \arrow[u]                                                                     & 0 \arrow[u]                                                                     &                            & 0 \arrow[u]                                                               &  
\end{tikzcd}
\ee
Here each $C^p(\mathfrak{U},\Omega^q_M)$ denotes the space of $p$-cochains of the cover $\mathfrak{U}$ with coefficients in $q$-forms on $M$, such that their elements $\varphi_{i_0 i_1 \ldots i_p}$ are defined on the intersection $U_{i_0} \cap U_{i_1} \cap \cdots \cap U_{i_p}$, see, e.g., \cite{Bott1982} for a textbook reference. For instance, in the two extreme cases $p=0$ and $p=m{-}1$, which will be of our main interest, we have
\be
C^0(\mathfrak{U},\Omega^q_M) = \bigoplus_{i=0}^{m-1} \Omega^q_{U_i}, \qquad
C^{m-1}(\mathfrak{U},\Omega^q_M) = \Omega^q_{M-\Crit(W)}.
\ee
Each vertical line in \eqref{Cech-de-Rham} then becomes copies of a twisted de Rham complex with a differential $\nabla_{dW}$. In the horizontal direction we have a {\v C}ech coboundary operator $\delta$ satisfying $\delta^2 = 0$, which acts as
\be
(\delta \varphi)_{i_0 i_1 \ldots i_{p+1}} = \sum_{r=0}^{p+1} (-1)^r \varphi_{i_0 \ldots \widehat{i}_r \ldots i_{p+1}},
\ee
where the hat denotes an omitted index. One can check that $\delta\, \nabla_{dW} - \nabla_{dW}\, \delta = 0$.
Let us group terms along the anti-diagonal of \eqref{Cech-de-Rham} by defining
\be
K^r := \bigoplus_{p+q = r} C^p(\mathfrak{U}, \Omega^q_M),
\ee
followed by an introduction of the differential operator $D: K^{p+q} \to K^{p+q+1}$ given by
\be
D := \delta + (-1)^p \nabla_{dW}.
\ee
One can check that it satisfies
\be
D^2 = \delta^2 + \delta\, \nabla_{dW} - \nabla_{dW}\, \delta + \nabla_{dW}^2  = 0,
\ee
which allows us to define a cohomology of the complex $(K^\bullet,D)$ often called the \emph{hypercohomology} of the cover $\mathfrak{U}$ with coefficients in the double complex \eqref{Cech-de-Rham} and denoted by $\mathbb{H}^p(\mathfrak{U},(\Omega^\bullet_M,\nabla_{dW}))$. As before, replacing $W \to -W$ at all steps allows us to define a dual hypercohomology.

The intersection number \eqref{intersection-definition} can be re-stated in this formulation as \cite{saito1983higher}
\be\label{alt-definition}
\la \varphi_- | \varphi_+ \ra_{dW} = \tau^m \Res_{dW=0} \left( \varphi_- \psi_+ \right),
\ee
where $\psi_+ \in C^{m-1}(\mathfrak{U},{\cal O}_M)$ is a {\v C}ech-dual function to $\varphi_+$. The equivalence to \eqref{intersection-definition} follows from the fact that both definitions satisfy Saito's uniqueness theorem \cite{saito1983higher}, up to an overall constant. This constant is fixed by matching the leading $\tau \to \infty$ asymptotics of the intersection number to \eqref{residue-pairing}, which was done independently in \cite{Mizera:2017rqa}, and determines the prefactor $\tau^m$ on the right-hand side of \eqref{alt-definition} in our conventions.

We can compute $\psi_+$ as follows. Let us first use the embedding
\be
\jmath: H^{m}(M,\nabla_{dW}) \,\to\, C^0(\mathfrak{U}, \Omega^m_M),
\ee
so that $\jmath(\varphi_+) \in C^0(\mathfrak{U},\Omega_M^m)$ defines an element in the top-left corner of the twisted {\v C}ech--de Rham complex \eqref{Cech-de-Rham}. Explicitly,
\be\label{j-varphi}
\jmath(\varphi_+) = \left( \widehat{\varphi}_+\, dz_1 \wedge \cdots \wedge \widehat{dz}_{i_0+1} \wedge \cdots \wedge dz_m \right)_{\!i_0 = 0,1,\ldots,m-1}.
\ee
Then $\psi_+$, in the bottom-right corner, can be obtained by solving $D \Psi_+ = \varphi_+$ and extracting $\psi_+$ as the $p=m{-}1$ component of $\Psi_+$. Concretely this can be done by tracing a zig-zag path through the diagram:
\be
\begin{tikzcd}
{C^0(\mathfrak{U},\Omega^m_M)}                                                     &                                                                                    &                                                        &                                                                    \\
{C^0(\mathfrak{U},\Omega^{m-1}_M)} \arrow[u, "\nabla_{dW}"'] \arrow[r, "\delta"] & {C^1(\mathfrak{U},\Omega^{m-1}_M)}                                                 &                                                        &                                                                    \\
& {C^1(\mathfrak{U},\Omega^{m-2}_M)} \arrow[u, "\nabla_{dW}"'] \arrow[r, "\delta"] & {C^2(\mathfrak{U},\Omega^{m-2}_M)}                     &                                                                    \\
&                                                                                    & \ddots \arrow[r, "\delta"] \arrow[u, "\nabla_{dW}"'] & {C^{m-1}(\mathfrak{U},\Omega^1_M)}                                 \\
&                                                                                    &                                                        & {C^{m-1}(\mathfrak{U},{\mathcal O}_M)} \arrow[u, "\nabla_{dW}"']
\end{tikzcd}
\ee
This gives us an expression for $\psi_+$, which involves applying the inverse operator $\nabla_{dW}^{-1}$ $m$ times and $\delta$ $m{-}1$ times in alternating order:
\be\label{psi-plus}
\psi_+ = \nabla^{-1}_{dW} (\delta \nabla_{dW}^{-1})^{m-1} \jmath(\varphi_+).
\ee
Since we are interested in explicit formulae, let us show how to compute $\psi_+$ step-by-step. It will be convenient to introduce the following notation for each component of $\nabla_{dW}$,
\be
\nabla_{dW} =: \sum_{i=1}^{m} \nabla_i\, dz_i,
\ee
so that $\nabla_i = \partial_i + \tau \partial_i W$.
Starting from \eqref{j-varphi}, let us spell out first couple of steps in evaluating \eqref{psi-plus}:
\begin{gather}
\nabla_{dW}^{-1} \jmath(\varphi_+) = \left( \nabla_{i_0+1}^{-1} \widehat{\varphi}_+\, dz_1 \wedge \cdots \wedge \widehat{dz}_{i_0+1} \wedge \cdots \wedge dz_m \right)_{\!i_0 = 0,1,\ldots,m-1},\\
\nabla_{dW}^{-1} \delta \nabla_{dW}^{-1} \jmath(\varphi_+) = \left( \nabla_{i_0+1}^{-1}\nabla_{i_1+1}^{-1} \widehat{\varphi}_+\, dz_1 \wedge \cdots \wedge \widehat{dz}_{i_0+1} \wedge \cdots \wedge \widehat{dz}_{i_1+1} \wedge \cdots \wedge dz_m \right)_{\!0 \leq i_0 < i_1 \leq m-1},\nn
\end{gather}
from which the general patter should be clear (the offset by $1$ is simply a consequence of our conventions for the covering $\mathfrak{U} = \{U_{i+1}\}_{i=0}^{m-1}$ whose index traditionally starts from $0$ instead of $1$). After $m$ steps we find
\be
\psi_+ = \nabla_{1}^{-1} \nabla_{2}^{-1} \cdots \nabla_{m}^{-1} \,  \widehat{\varphi}_+,
\ee
where the inverses $\nabla_i^{-1}$ are understood in terms of their expansion around the $\tau \to \infty$ limit. Note that $\nabla_{i}^{-1}$'s commute and hence we do not need to specify the order in which they are applied. We will evaluate each order in $\tau^{-1}$ in the following subsection.

As a matter of fact, we can derive a dual formula for intersection numbers given by
\be\label{alt-definition-2}
\la \varphi_- | \varphi_+ \ra_{dW} = (-\tau)^m \Res_{dW=0} \left( \psi_- \varphi_+ \right),
\ee
whose evaluation is entirely analogous. Defining $\nabla_{-dW} =: \sum_{i=1}^{m} \nabla_{-i}\, dz_i$ one finds
\be
\psi_- = \nabla_{-1}^{-1} \nabla_{-2}^{-1} \cdots \nabla_{-m}^{-1} \,  \widehat{\varphi}_-.
\ee
In general, one can interpolate between the two definitions \eqref{alt-definition} and \eqref{alt-definition-2} by considering a paring 
\be
C^{p}(\mathfrak{U},\Omega_M^{q}) \times C^{r}(\mathfrak{U},\Omega_M^{s}) \,\to\, \C
\ee
with $p{+}r=m{-}1$ and $q{+}s=m$, however we will not do so here.

\subsection{\label{sec:higher-residue-pairings}Higher Residue Pairings}

To summarize, we found that intersection numbers of the cohomology classes $[\varphi_\pm] \in H^{m}_{\pm dW}$ can be expressed in terms of Grothendieck residues around the critical points $\Crit(W)$ as follows:
\begin{align}\label{formula1}
\la \varphi_- | \varphi_+ \ra_{dW} &= \tau^m \Res_{dW=0} \Big( \widehat{\varphi}_- \nabla_{1}^{-1} \nabla_{2}^{-1} \dots \nabla_{m}^{-1} \,  \widehat{\varphi}_+\, d^m z \Big),
\end{align}
where $\nabla_i = \partial_i + \tau \partial_i W$, $\varphi_\pm = \widehat{\varphi}_\pm d^mz$, and the residue is defined as in \eqref{residue-def}. We also have a dual formula for the same object, obtained by switching the roles of $\varphi_-$ and $\varphi_+$:
\be\label{formula2}
\la \varphi_- | \varphi_+ \ra_{dW} = (-\tau)^m \Res_{dW=0} \Big( \widehat{\varphi}_+ \nabla_{-1}^{-1} \nabla_{-2}^{-1} \dots \nabla_{-m}^{-1} \,  \widehat{\varphi}_-\, d^m z \Big),
\ee
where $\nabla_{-i} = \partial_i - \tau \partial_i W$. We are interested in expanding such intersection numbers in powers of $\tau^{-1}$, as follows:
\be
\la \varphi_- | \varphi_+\ra_{dW} =: \sum_{k=0}^{\infty} \tau^{-k} (\varphi_- | \varphi_+)_{dW,k},
\ee
where the leading order starts at $\tau^0$ because of the overall normalization of intersection numbers. The coefficients $(\varphi_- | \varphi_+)_{dW,k}$ are called \emph{higher residue pairings} \cite{saito1983higher}. They have the symmetry property
\be
( \varphi_- | \varphi_+ )_{dW,k} = (-1)^{k} (\varphi_+ | \varphi_- )_{dW,k}.
\ee
For instance, when $\varphi_+ = \varphi_-$ all the odd higher residue pairings vanish identically.

In the following subsections we extract higher residue pairings directly from the expressions \eqref{formula1} and \eqref{formula2} by expanding each inverse derivative operator according to
\be\label{nabla-inverse-expansion}
\nabla^{-1}_{i} \eta = \tau^{-1} \frac{\eta}{\partial_i W} - \tau^{-2} \frac{1}{\partial_i W} \partial_i \left( \frac{\eta}{\partial_i W} \right) + \tau^{-3} \frac{1}{\partial_i W} \partial_i \left( \frac{1}{\partial_i W} \partial_i \left( \frac{\eta}{\partial_i W} \right) \right) - \dots,
\ee
which can be shown by requiring that $\nabla_i \nabla_i^{-1} \eta = \eta$ order-by-order. Likewise we have
\be
\nabla^{-1}_{-i} \eta = -\tau^{-1} \frac{\eta}{\partial_i W} - \tau^{-2} \frac{1}{\partial_i W} \partial_i \left( \frac{\eta}{\partial_i W} \right) - \tau^{-3} \frac{1}{\partial_i W} \partial_i \left( \frac{1}{\partial_i W} \partial_i \left( \frac{\eta}{\partial_i W} \right) \right) - \dots,
\ee
which is obtained simply by replacing $\tau \to - \tau$ in \eqref{nabla-inverse-expansion}.
In general the two types of expansions will involve similarly-looking terms that could cancel out upon averaging between \eqref{formula1} and \eqref{formula2}. We exploit this fact in deriving explicit expressions for $k=0,1,2$ below.

\subsubsection{Leading Order}

At the leading order we see straightforwardly that the two expression evaluate to
\be\label{leading-order}
( \varphi_- | \varphi_+ )_{dW,0} = \Res_{dW=0} \left( \frac{\widehat{\varphi}_- \widehat{\varphi}_+ \, d^m z}{\partial_1 W\, \partial_2 W\, \cdots\, \partial_m W} \right),
\ee
which was in fact shown previously in \cite{Mizera:2017rqa} using complex Morse theory. In order to evaluate \eqref{leading-order} explicitly, let us make use of the $m\times m$ Hessian matrix $\mathbf{\Phi}$ with entries
\be
\mathbf{\Phi}_{ij} := \partial_i \partial_j W,
\ee
which, by the assumption on non-degeneracy of the critical points, has maximal rank and hence is invertible. To compute the residue we first change the form variables from $(z_1,z_2,\ldots, z_m)$ to $(\partial_1 W, \partial_2 W, \ldots, \partial_m W)$, at a cost of dividing by the Jacobian $\det \mathbf{\Phi}$, so that we obtain
\begin{align}
( \varphi_- | \varphi_+ )_{dW,0} &= \Res_{dW=0} \left( \frac{1}{\det \mathbf{\Phi}} \frac{\widehat{\varphi}_- \widehat{\varphi}_+ \, d(\partial_1 W) \wedge d(\partial_2 W) \wedge \cdots \wedge d(\partial_m W) }{\partial_1 W\, \partial_2 W\, \cdots\, \partial_m W} \right) \nn\\
&= \sum_{(z_1^\ast, z_2^\ast, \ldots, z_m^\ast) \in \Crit(W)} \frac{\widehat{\varphi}_- \widehat{\varphi}_+}{\det \mathbf{\Phi}} \Bigg|_{z_i = z_i^\ast}.
\end{align}
Here we also used the fact that $\widehat{\varphi}_\pm$ do not have poles on the critical locus $\Crit(W)$.

It is known that if both twisted forms $\varphi_\pm$ are logarithmic, their intersection number is homogeneous in $\tau$ and $(\varphi_- | \varphi_+)_{dW,0}$ is the only non-vanishing residue pairing, see, e.g., \cite{matsumoto1998,Mizera:2019gea}.

\subsubsection{Subleading Order}

Expanding \eqref{formula1} to order $\tau^{-1}$ we find at subleading order
\begin{align}
( \varphi_- | \varphi_+ )_{dW,1} &= \Res_{dW=0} \left(  \frac{\widehat{\varphi}_-\, d^mz}{\partial_1 W \partial_2 W \cdots \partial_m W} \sum_{i=1}^{m} \left( -\frac{\partial_i \widehat{\varphi}_+}{\partial_i W} + \frac{\widehat{\varphi}_+}{\partial_i W} \sum_{j=i}^{m} \frac{\partial_i \partial_j W}{\partial_j W} \right) \right) \nn \\
&= \Res_{dW=0} \left( \!{-}\!\sum_{i=1}^{m} \frac{\widehat{\varphi}_- \partial_i \widehat{\varphi}_+\, d^mz}{\partial_1 W \cdots (\partial_i W)^2 \cdots \partial_m W} + \sum_{i=1}^{m} \frac{\widehat{\varphi}_- (\partial^2_i W) \widehat{\varphi}_+\, d^mz}{\partial_1 W \cdots (\partial_i W)^3 \cdots \partial_m W} \right. \nn \\
&\qquad\qquad\qquad\qquad\qquad \left. + \sum_{i < j} \frac{\widehat{\varphi}_- (\partial_i \partial_j W) \widehat{\varphi}_+\, d^mz}{\partial_1 W \cdots (\partial_i W)^2 \cdots (\partial_j W)^2 \cdots \partial_m W} \right).
\end{align}
Using the expression \eqref{formula2} leads to a similar expression which is related to the above one by exchanging $\varphi_- \leftrightarrow \varphi_+$ and an overall minus sign. Thus after symmetrizing the result the final two terms cancel out and we are left with
\be
( \varphi_- | \varphi_+ )_{dW,1} = \frac{1}{2} \Res_{dW=0} \left( \sum_{i=1}^{m} \frac{ \left( \widehat{\varphi}_+ \partial_i \widehat{\varphi}_- - \widehat{\varphi}_- \partial_i \widehat{\varphi}_+ \right) d^mz}{\partial_1 W \cdots (\partial_i W)^2 \cdots \partial_m W} \right),
\ee
which matches the subleading higher residue pairing \cite{saito1983higher}.

Repeating the steps from the previous subsection we obtain:
\begin{align}
( \varphi_- | \varphi_+ )_{dW,1} &=
\frac{1}{2} \sum_{(z_1^\ast, z_2^\ast, \ldots, z_m^\ast) \in \Crit(W)}  \sum_{i=1}^{m} \frac{d}{d(\partial_i W)} \left( \frac{ \widehat{\varphi}_+ \partial_i \widehat{\varphi}_- - \widehat{\varphi}_- \partial_i \widehat{\varphi}_+}{\det \mathbf{\Phi}}\right) \Bigg|_{z_k = z_k^\ast} \nn\\
&= \frac{1}{2} \sum_{(z_1^\ast, z_2^\ast, \ldots, z_m^\ast) \in \Crit(W)}  \sum_{i,j=1}^{m} \mathbf{\Phi}^{-1}_{ij} \frac{\partial}{\partial z_j} \left( \frac{ \widehat{\varphi}_+ \partial_i \widehat{\varphi}_- - \widehat{\varphi}_- \partial_i \widehat{\varphi}_+}{\det \mathbf{\Phi}}\right) \Bigg|_{z_k = z_k^\ast},
\end{align}
where the first line is a result of performing a residue around the double pole in $\partial_i W$, while in the second line we changed the variables back to $(z_1, z_2, \ldots, z_m)$ in order to evaluate the derivative explicitly.

\subsubsection{Subsubleading Order}

Expanding \eqref{formula1} to order $\tau^{-2}$, we find that the subsubleading correction to the intersection number is given by the higher residue pairing
\begin{align} \label{subsubleading}
(\vphi_- \vert \vphi_+)_{dW,2} &= \Res_{dW=0}
\!\Bigg( 
    \widehat\vphi_{-}
    \Bigg(
    \sum_{i=1}^m \frac{1}{\prod_{k=1}^i\partial_kW}
    \partial_i \left(
        \frac{1}{\partial_iW} \partial_i 
        \left(
            \frac{\widehat\vphi_+}{\prod_{l=i}^m\partial_lW}
        \right)
    \right)
 \\
& \qquad\qquad\qquad
    + \sum_{i=1}^{m} \sum_{j=i+1}^m 
    \frac{1}{\prod_{k=1}^i\partial_kW}
    \partial_i
    \bigg(
        \frac{1}{\prod_{l=i}^j\partial_lW}
        \partial_j 
        \bigg(
            \frac{\widehat\vphi_+}{\prod_{p=j}^m\partial_pW}
        \bigg)
    \bigg)
    \Bigg)
    d^mz\!
\Bigg).\nn
\end{align}
Due to the complicated combinatorics we will not explicitly expand the derivatives above. However, let us comment on some features of the above expression. The subsubleading term has poles in $\partial_iW$ of maximal order $5$, which come from the first line of equation \eqref{subsubleading}. After expanding the derivatives one finds multiple sums, the largest being a $4$-fold sum with ${\cal O}(m^4)$ terms. Equation \eqref{formula2} also leads to a similar expression for the subsubleading pairing, which is related to \eqref{subsubleading} by exchanging $\varphi_- \leftrightarrow \varphi_+$, and hence there would be no cancellations after symmetrizing.

\subsection{\label{sec:integration}Twisted Periods}

Before closing this section let us explain where the interest in intersection numbers comes from in the present context. As remarked before, many quantities of physical interest can be written as integrals on $M$ of the general form:
\be\label{general-integral}
\int_{\Gamma} e^{\tau W} \varphi,
\ee
for a multi-valued function $W \in {\cal O}_{\widehat{M}}$, a middle-dimensional cycle $\Gamma \subset M$, and a single-valued form $\varphi \in \Omega^m_M$. We assume that the pole divisor of $\varphi$ and boundaries of $\Gamma$ are contained in the divisor of $M$. Such integrals have a natural interpretation as bilinear pairings
\be
H_m^{dW} \times H^m_{dW} \to \C,
\ee
between elements of twisted cohomology groups $[\varphi] \in H^m_{dW}$ and (locally-finite) twisted homology groups $[\Gamma] \in H_m^{dW}$. Therefore we will refer to integrals of the form \eqref{general-integral} as \emph{twisted periods}. The precise definition of the $H_m^{dW}$ does not matter for our purposes (see, e.g., Appendix~A of \cite{Mizera:2019gea} for an exposition), other than the fact that it leads to the above pairing \cite{aomoto2011theory}. For physical applications see, e.g., \cite{varchenko1990multidimensional,doi:10.1142/2467,Mimachi2003,Mimachi2004,alex2004bethe,Schwarz:2008sa,Varchenko_2011,Mizera:2017cqs,Li:2018mnq,Mastrolia:2018uzb,Frellesvig:2019kgj,Mizera:2019gea,Frellesvig:2019uqt,Brown:2019wna,Abreu:2019wzk,Casali:2019ihm,CHP} and references therein.

Given that the dimension of $H^m_{dW}$ is $|\chi(M)|$, an arbitrary twisted form $\varphi$ can be expanded into a basis $\{\varphi_a\}_{a=1}^{|\chi(M)|}$ of this cohomology group. This can be done explicitly by 
\be\label{basis-expansion}
\varphi = \sum_{a=1}^{|\chi(M)|} \la \varphi_a^\vee | \varphi \ra_{dW} \; \varphi_a,
\ee
where $\{\varphi_a^\vee\}_{a=1}^{|\chi(M)|}$ is a basis of the dual cohomology group $H^{m}_{-dW}$, which is orthonormal in the sense that $\la \varphi_a^\vee | \varphi_b \ra_{dW} = \delta_{ab}$. Naturally, the above equality implies a relation between integrals \eqref{general-integral}. As a matter of fact, similar decomposition can be achieved in the homology basis, leading to a $|\chi(M)| \times |\chi(M)|$ period basis of integrals, but we do not use it as in our applications $\Gamma$ is always kept constant.

There is one caveat, however, in that an orthonormal set of bases might not be easily found, as is in fact generically the case. This can be alleviated by introducing an auxiliary basis $\{\vartheta_b\}_{b=1}^{|\chi(M)|}$ of $H^{m}_{-dW}$ to write down
\be\label{non-ortho}
\la \varphi_a^\vee | \varphi \ra_{dW} = \sum_{b=1}^{|\chi(M)|} \,\mathbf{C}^{-1}_{ab}\, \la \vartheta_{b} | \varphi \ra_{dW} \qquad\mathrm{with}\qquad \mathbf{C}_{ba} := \la \vartheta_b | \varphi_a \ra_{dW},
\ee
which follows from a simple linear algebra exercise. In this way we can use \eqref{basis-expansion} to perform expansion of an arbitrary integral into a basis. This simple property, when used together with higher residue pairings, turns out to be quite powerful.

\section{\label{sec:from-infinite-to-four-dimensions}From Infinity to Four Dimensions}

In this section we apply the formalism reviewed above to extract the information about analytic properties of Feynman integrals. After reviewing their representation as twisted periods in Section~\ref{sec:Feynman-integrals}, we discuss how to construct vector bundles of such integrals over the kinematic space in Section~\ref{sec:vector-bundles}, followed by a determination of their connections in terms of intersection numbers and higher residue pairings. We finish with explicit examples in Sections~\ref{sec:massless-box}--\ref{sec:massive-sunrise} for one- and two-loop diagrams.

\subsection{\label{sec:Feynman-integrals}Feynman Integrals as Twisted Periods}

Let us start by reviewing how to translate a given Feynman integral from its momentum-space form into a representation using Schwinger parameters. An $L$-loop integral with $P$ propagators $\{{\mathsf D}_a\}_{a=1}^{P}$ is given by
\be\label{Feynman-integral}
I_{\nu_1,\nu_2,\ldots,\nu_P} := \frac{1}{(i\pi^{D/2})^L} \int  \frac{\prod_{i=1}^{L} d^D \ell_i}{\prod_{a=1}^{P} \mathsf{D}_a^{\nu_a}},
\ee
where the integration contour is $(\ell_1, \ell_2,\ldots,\ell_L) \in (\R^{1,D-1})^L$ in Lorentzian signature and the overall constant is for later convenience. We use mostly-plus conventions for the metric. Each integral is labelled by integers $(\nu_1, \nu_2, \ldots, \nu_P) \in \Z^P$ that specify the powers to which the corresponding denominators are taken. There is no substantial difficulty in repeating our analysis for multi-loop scattering amplitudes (sums over multiple Feynman integrals of the above type), by allowing the set $\{\mathsf{D}_a\}_{a=1}^{P}$ to be large enough, however we focus on individual integrals in order to make universal statements that do not depend on a specific quantum field theory.

We employ a ``Schwinger trick'' in which each denominator ${\mathsf D}_a$ is expressed as an integral over a variable $x_a$ representing Schwinger time associated to the corresponding edge of the Feynman diagram:
\be\label{Schwinger-trick}
\frac{1}{\mathsf{D}_a^{\nu_a}} = \frac{1}{\Gamma(\nu_a)} \int_{\R_+} x_a^{\nu_a - 1} e^{-x_a {\mathsf D}_a} dx_a.
\ee
For the time being let us not worry about a possible divergence of the gamma function and treat $\nu_a$ as formal parameters. Applying \eqref{Schwinger-trick} to the above Feynman integral $P$ times will involve the following combination in the exponent:
\be
\sum_{a=1}^{P} x_a {\mathsf D}_a =: \sum_{i,j=1}^{L} {\mathbf Q}_{ij}\, \ell_i {\cdot} \ell_j + 2 \sum_{i=1}^{L} {\vec L}_i {\cdot} \ell_i + c,
\ee
which defines the $L\times L$ matrix ${\mathbf Q}$, the length-$L$ vector ${\vec L}$, and the scalar $c$ in terms of kinematic invariants and $x_a$'s. Since our goal is to integrate out the loop momenta, we first complete the square in this combination:
\be
\sum_{a=1}^{P} x_a {\mathsf D}_a = (\ell + {\mathbf Q}^{-1} {\vec L})^{\intercal} {\mathbf Q} (\ell + {\mathbf Q}^{-1} {\vec L}) + c - {\vec L}^\intercal {\mathbf Q}^{-1} {\vec L}.
\ee
The Gaussian integral over $\ell_i$'s gives $(i\pi^{D/2})^L/(\det {\mathbf Q})^{D/2}$ thus cancelling the prefactor in \eqref{Feynman-integral} (the factors of $i$ come from a Wick rotation to Euclidean time) and the resulting expression becomes
\be\label{parametric}
I_{\nu_1,\nu_2,\ldots,\nu_P} = \frac{1}{\prod_{a=1}^{P} \Gamma(\nu_a)} \int_{\R_+^P} \frac{e^{{\vec L}^\intercal {\mathbf Q}^{-1} {\vec L} - c}}{(\det {\mathbf Q})^{D/2}} \prod_{a=1}^{P} x_a^{\nu_a - 1} dx_a.
\ee
The integration contour is now independent of the space-time dimension $D$ and from now on we employ dimensional regularization by setting $D = 4-2\varepsilon$.

Let us briefly comment on the meaning of the space $M$ with coordinates $(x_1, x_2, \ldots, x_P)$. Given a graph $G$ whose internal edges are prescribed by the set of propagators $\{\mathsf{D}_a\}_{a=1}^{P}$, each Schwinger parameter $x_a$ parametrizes proper length the edge associated to ${\mathsf D}_a$. For this reason we will refer to $M$ as the \emph{moduli space} ${\cal M}_G$ of Riemannian metrics on $G$.

The above integral is already in a form similar to \eqref{general-integral}, however for our purposes we would like to massage it into an integral where the potential $W$ is proportional to $\varepsilon$, i.e., identify it with the expansion parameter $\tau$ in \eqref{general-integral}. To this end we follow a standard procedure by inserting $1 = \int_{\R_+} \!\delta(\rho - \sum_{a=1}^{M} x_a) d\rho$ into \eqref{parametric}, followed by rescaling $x_a \to \rho x_a$. Collecting all the Jacobians this leaves us with
\be
I_{\nu_1,\nu_2,\ldots,\nu_P} = \frac{1}{\prod_{a=1}^{P} \Gamma(\nu_a)} \int_{\R_+^{P+1}} \rho^{|\nu| + L(\varepsilon-2) -1} \frac{e^{-\rho {\cal F}/{\cal U}}}{{\cal U}^{2-\varepsilon}} d\rho\; \delta(1- {\textstyle\sum_{a=1}^{P}} x_a) \prod_{a=1}^{P} x_a^{\nu_a - 1} dx_a,
\ee
where $|\nu| := \sum_{a=1}^{P} \nu_a$. For brevity of notation we also expressed the result in terms of the so-called Symanzik polynomials:
\be\label{Symanzik-polynomials}
{\cal U} := \det \mathbf{Q}, \qquad {\cal F} := {\cal U} (c - {\vec L}^\intercal {\mathbf Q}^{-1} {\vec L}).
\ee
One can recognize that the $\rho$ integral is of the form \eqref{Schwinger-trick} and hence evaluates to
\be\label{rho-integral}
\int_{\R_+} \!\!\rho^{|\nu| + L(\varepsilon-2)-1} \frac{e^{-\rho {\cal F}/{\cal U}}}{{\cal U}^{2-\varepsilon}} d\rho =  \frac{\Gamma(|\nu|{+}L(\varepsilon{-}2))}{{\cal U}^{2-\varepsilon} \left({\cal F}/{\cal U}\right)^{|\nu|+L(\varepsilon-2)}}.
\ee
On the other hand we can rewrite the right-hand side of \eqref{rho-integral} using Feynman parametrization as
\be
\frac{\Gamma(|\nu|{+}L(\varepsilon{-}2))}{{\cal U}^{(L+1)(2-\varepsilon)-|\nu|} {\cal F}^{|\nu|+L(\varepsilon-2)}} = \frac{\Gamma(2{-}\varepsilon)}{\Gamma((L{+}1)(2{-}\varepsilon){-}|\nu|)} \int_{\R_+} \frac{\tilde{\rho}^{|\nu|+L(\varepsilon-2)-1} d\tilde{\rho}}{({\cal U}+\tilde{\rho} {\cal F})^{2-\varepsilon}}.
\ee
Followed by rescaling $x_a \to x_a/\tilde{\rho}$ and undoing the $\tilde\rho$ integration with $\int_{\R_+} \delta(\tilde\rho - \sum_{a=1}^{M} x_a) d\tilde\rho = 1$ we obtain the representation \cite{Lee:2013hzt}:
\be\label{parametric-representation}
I_{\nu_1,\nu_2,\ldots,\nu_P} = \frac{\Gamma(2{-}\varepsilon)}{\Gamma((L{+}1)(2{-}\varepsilon){-}|\nu|)\prod_{a=1}^{P} \Gamma(\nu_a)} \int_{\R_+^P} ({\cal F}{+}{\cal U})^{\varepsilon-2} \prod_{a=1}^{P} x_a^{\nu_a - 1} dx_a.
\ee

This family of integrals is almost of the form \eqref{general-integral}, if it were not for the following fact. Taking the potential function suggested by the above representation, $\tau W = \varepsilon \log({\cal F}+{\cal U})$, leaves us with forms $\varphi$ of the type $\prod_{a=1}^{M} x_a^{\nu_a - 1} dx_a/({\cal F}{+}{\cal U})^2$. These might have poles on $\{x_a = 0\}$ and/or $\{x_a = \infty\}$ depending on the values of $\nu_a$'s, which violate the assumption that the pole divisor of $\varphi$ is contained within the divisor of the integration space $M$ (similar issue appears in the definition of the integration cycle $\R_+^P$). This is actually a physical effect, since such singularities of $\varphi$ correspond to propagators pinching, and thus cannot be removed. It signals that one should have instead considered a twisted homology \emph{relative} to such singularities, see \cite{CHP} for a formulation of Feynman integrals in this setup.

Nevertheless, the philosophy of the present paper is that one should study properties of integrals on $M$ \emph{globally} and in particular without worrying about stratification of its boundaries and related issues. Thus we follow a different path and redefine \eqref{parametric-representation} by infinitesimally deforming the integer parameters $\nu_a$ to
\be
\nu_a \to \nu_a + \varepsilon \delta_a,
\ee
where each $\delta_a$ is a generic infinitesimal variable. We send each $\delta_a$ to zero at the end of the computation and assume that the resulting regulated integrals $\widehat{I}_{\nu_1,\nu_2,\ldots,\nu_P}$ are smooth in this limit. With this deformation we have
\be\label{superpotential}
W = \log({\cal F}{+}{\cal U}) + \sum_{a=1}^{P} \delta_a \log (x_a),
\ee
and since $M$ is defined as $\CP^m$ minus the pole divisor of $dW$, boundaries of $M$ now include $\{x_a = 0\}$ and $\{x_a = \infty\}$ for all $a$. We define twisted cohomologies $H^P_{\pm dW}$ with this potential and identify $\tau = \varepsilon$. This leads to a family of twisted forms:
\be
\varphi_{\nu_1, \nu_2, \ldots, \nu_P} := \frac{\Gamma(2{-}\varepsilon)}{\Gamma((L{+}1)(2{-}\varepsilon){-}|\nu|{-}\varepsilon|\delta|)\prod_{a=1}^{P} \Gamma(\nu_a{+}\varepsilon\delta_a)} ({\cal F}{+}{\cal U})^{-2} \bigwedge_{a=1}^{P} x_a^{\nu_a - 1} dx_a,
\ee
where $|\delta|:= \sum_{a=1}^{P} \delta_a$. We will often set all $\delta_a$'s to be equal, $\delta_a = \delta$. Note that the deformation also regularized the gamma functions. In this language (regulated) Feynman integrals become twisted periods:
\be\label{I-hat}
\widehat{I}_{\nu_1, \nu_2, \ldots, \nu_P} := \int_{\Gamma} e^{\varepsilon W} \varphi_{\nu_1, \nu_2, \ldots, \nu_P},
\ee
where the middle-dimensional integration cycle is $\Gamma := \R_+^{P}$ and the hat denotes the fact that we expect \eqref{I-hat} to agree with \eqref{Feynman-integral} only after taking the limit $\delta_a \to 0$. Closely related ways of rewriting Feynman integrals as twisted periods were introduced in \cite{Mastrolia:2018uzb}, see also \cite{Frellesvig:2019kgj,MastroliaTalk,delaCruz:2019skx,Frellesvig:2019uqt,sameshima2019different,Abreu:2019wzk,Klausen:2019hrg,CHP}.

To simplify $\varepsilon$-power counting we will normalize basis forms by appropriate powers of $\varepsilon$ such that they start at $\varepsilon^0$. In addition, rescaling the integration variables $x_a \to \beta z_a$ by a constant $\beta$ typically allows one to remove one mass-scale outside of the integral, which is what we will do in the explicit examples below.

The dimension of twisted cohomology groups $H^{P}_{\pm dW}$ is the absolute value of the Euler characteristic $|\chi(M)|$ of $M$, which is given by
\be
M := (\C^\times)^P - \{ {\cal F}{+}{\cal U} = 0 \},
\ee
where $\C^\times := \C {-} \{0\}$, in agreement with \cite{Bitoun:2017nre,Frellesvig:2019uqt}. Physically it counts the number of linearly-independent Feynman integrals that involve the set of propagators $\{{\mathsf D}_a\}_{a=1}^{P}$ over $\Q(K,\varepsilon,\delta_a)$, where $K$ in the set of kinematic variables appearing in ${\mathbf Q}, {\mathbf L}, c$.\footnote{In the Feynman integral literature it is conventional to call elements of the basis ``master integrals''. We prefer not to use such nomenclature due to its many ambiguities stemming from distinct definitions of the same term being used by different authors. Basis of twisted cohomology, as defined presently, corresponds to Feynman integrals (reducible and irreducible) in all sectors, without imposing any additional non-linear symmetries.} It is the most convenient to compute $|\chi(M)|$ by invoking Morse-theory arguments, which for sufficiently generic $W$ imply that it is equal to the number of critical points $\Crit(W)$ determined by the condition $dW=0$. From this point of view the regulators $\delta_a$ are needed to ensure that the Morse flow is transverse to the divisors $\{x_a = 0\}$ and $\{x_a = \infty \}$. Explicitly, $dW=0$ gives a system of equations
\be\label{critical-points}
\frac{\partial_a ({\cal F}{+}{\cal U})}{{\cal F}{+}{\cal U}} + \frac{\delta_a}{x_a} = 0
\ee
for $a=1,2,\ldots,m$. Recall that in our work we always assume the critical points are isolated and non-degenerate. Since we will be interested only the limit $\delta_a \to 0$, it is sufficient to solve the above constraints as an expansion of $z_a$ in $\delta_a$, which greatly simplifies finding the critical points. (One should \emph{not} expand the equations \eqref{critical-points} themselves, as the number of solutions is generically discontinuous in such a procedure.) According to \cite{Lee:2013hzt}, the number of critical points that solve \eqref{critical-points} with strictly $\delta_a=0$ computes the number of top-level Feynman integrals in a given family.

There is no agreed-upon way of finding bases of Feynman integrals. There \emph{is}, however, a criterion for what constitutes a ``good'' basis based on its behaviour near the $\varepsilon \to 0$ limit. This leads us to the following discussion.

\subsection{\label{sec:vector-bundles}Vector Bundles Over the Kinematic Space}

Feynman integrals are functions on the kinematic space ${\cal K}$ with coordinates given by the kinematic variables $K$, such as Mandelstam variables or particles' masses. For our purposes, however, it is more convenient to think of a basis of Feynman integrals as a section of a vector bundle ${\cal V}$ over the kinematic space. To make this concrete, let us split the differential operator on the total space as $d = {\cal D} + \sum_{a=1}^{m} dz_a \partial_a$, where ${\cal D}$ acts only in the directions of the kinematic space. For a basis of twisted forms $\{\varphi_a\}_{a=1}^{|\chi(M)|} \in H^{P}_{dW}$ we have
\be
{\cal D} \int_{\Gamma} e^{\varepsilon W}\, \varphi_a \;=\; \sum_{b=1}^{|\chi(M)|} {\mathbf \Omega}_{ab} \int_{\Gamma} e^{\varepsilon W} \varphi_b,
\ee
where the matrix-valued one-form ${\bf \Omega}$ is given by intersection numbers \cite{Mastrolia:2018uzb}, as a special case of \eqref{basis-expansion}:
\be\label{Omega-definition}
{\mathbf \Omega}_{ab} := \la \varphi_{b}^{\vee} \,|\, ({\cal D}{+}\varepsilon {\cal D} W \wedge) \varphi_a \ra_{dW}.
\ee
Thus we can describe a basis of Feynman integrals $\vec{I} = \{I_a\}_{a=1}^{|\chi(M)|}$ as a section of ${\cal V}$, i.e., being defined by
\be\label{section}
({\cal D}-\mathbf{\Omega}\wedge) \vec{I} = 0.
\ee
Since the integration domain $\Gamma$ is always kept constant, each fiber of ${\cal V}$ is isomorphic to a twisted cohomology $H^{P}_{dW}$, where $W$ is determined by a point $K \in {\cal K}$ on the base space. Connections obtained in the above way are always integrable (flat), meaning that ${\cal D}-\mathbf{\Omega}\wedge$ squares to zero, i.e.,
\be\label{flatness}
{\cal D}\mathbf{\Omega}-\mathbf{\Omega}\wedge \mathbf{\Omega} = 0.
\ee
In physics parlance we are considering a non-abelian gauge theory, with zero curvature and gauge group  $\GL(|\chi(M)|,\C)$, on the kinematic space ${\cal K}$. Fixing a gauge corresponds to choosing a basis of Feynman integrals.

The equation \eqref{section}, together with a specification of boundary conditions, can be understood as an alternative definition of a family of Feynman integrals \cite{Kotikov:1990kg,Remiddi:1997ny,Gehrmann:1999as}. As a matter of fact, solving such differential equations provides one of the most efficient ways of evaluating Feynman integrals in practice, see, e.g., \cite{Smirnov:2006ry,Argeri:2007up,Henn:2014qga} for reviews. Our goal will therefore be to derive the connection in \eqref{section}.

In physically relevant situations the matrix ${\bf\Omega}$ can be expanded as a \emph{polynomial} in $\varepsilon$ \cite{Henn:2014qga},
\be\label{Omega-matrix}
{\bf \Omega} =: \sum_{k=0}^{k_{\text{max}}}  \varepsilon^k\, {\bf \Omega}_{(k)}
\ee
for $k_{\text{max}} < \infty$. To be more precise, if $\mathbf{\Omega}$ had any pole in $\varepsilon$, it would have to be spurious \cite{Henn:2014qga} and here we assume that such poles do not appear. This means we can evaluate intersection numbers that comprise the entries of $\mathbf\Omega$ either as an expansion around $\varepsilon \to 0$ or $\varepsilon \to \infty$ and both of them truncate. The latter expansion can be consistently carried out using higher residue pairings. As demonstrated in \cite{Mastrolia:2018uzb,Frellesvig:2019uqt,Frellesvig:2019kgj} the intersection numbers in $\mathbf\Omega$ can be also computed exactly in $\varepsilon$, though the techniques used there rely on the knowledge of either stratification of $M$ or its fibration properties. Using higher residue pairings allows us to forget about these technicalities.

In many cases one can bring \eqref{Omega-matrix} into a so-called \emph{$\varepsilon$-form}, where only a single term ${\bf \Omega}_{(1)}$ is non-vanishing \cite{Henn:2013pwa}. In those cases iterating the system of differential equations becomes particularly simple (an additional simplification would be if the matrix was triangular). A basis leading to an $\varepsilon$-form of $\mathbf\Omega$ is called \emph{canonical}. Literature on these aspects of differential equations includes \cite{Henn:2013pwa,Henn:2013fah,Henn:2013woa,Henn:2013nsa,barkatou2014reduction,Caron-Huot:2014lda,Argeri:2014qva,Gehrmann:2014bfa,Hoschele:2014qsa,Lee:2014ioa,Gituliar:2017vzm,Lee:2017qql,Herrmann:2019upk}.

There are two points we should discuss before diving into explicit computations. Firstly, the twisted form $({\cal D} + \varepsilon {\cal D}W\wedge) \varphi_a$ in \eqref{Omega-definition} is manifestly non-homogeneous in $\varepsilon$, and therefore we should consider the two terms ${\cal D}\varphi_a$ and $\varepsilon {\cal D}W\wedge\varphi_a$ separately to be consistent with counting powers of $\varepsilon$. Secondly, as remarked before the dual orthonormal basis $\{\varphi_b^\vee\}$ is typically not accessible a priori. Instead, we can use a more complicated expression
\be
{\mathbf \Omega}_{ab} = \sum_{c=1}^{|\chi(M)|} {\mathbf C}_{bc}^{-1} {\mathbf D}_{ca},
\ee
where $\mathbf{C}_{bc} := \la \vartheta_b | \varphi_c \ra_{dW}$ as in \eqref{non-ortho} and
\be\label{D-def}
{\bf D}_{ca} := \la \vartheta_c | {\cal D} \varphi_a \ra_{dW} + \varepsilon \la \vartheta_c | {\cal D}W {\wedge} \varphi_a \ra_{dW}.
\ee
Within the context of \emph{relative} twisted cohomology, one can systematically build orthonormal bases ($\mathbf{C}_{bc} = \delta_{bc}$) by choosing the dual forms to have compact support localized on the hypersurface defined by the set of propagators present in a given diagram \cite{CHP}.
Here we have introduced an auxiliary basis $\{ \vartheta_c \}_{c=1}^{|\chi(M)|}$ of $H_{-dW}^{P}$, which we take to be independent of $\vep$. In order to make $\varepsilon$-power counting simple, we normalize each $\varphi_a$ such that it contains only positive powers of $\vep$ up to the global maximum $\vep^n$
\begin{align}
    \vphi_a = \sum_{k=0}^{n} \vep^{k} \vphi_a^{(k)}
\end{align}
In terms of higher residue pairings we have simply:
\be\label{matrix-C}
{\bf C} =: \sum_{k=-n} \varepsilon^{-k} \, {\bf C}_{(-k)}
\quad \text{with} \quad 
{\bf C}^{(-k)}_{bc} = 
\begin{cases}
    \sum_{l=-n}^{k} 
    ( \vartheta_b | \varphi_c^{(-l)} )_{dW,k-l} 
    &\text{ for } k<0
    \\
    \sum_{l=0}^{n} 
    ( \vartheta_b | \varphi_c^{(l)} )_{dW,k+l} 
    &\text{ for } k \geq 0
\end{cases}
\ee
Taking into account non-homogeneity of twisted forms in $\mathbf{D}$ we have 
\be\label{matrix-D}
{\bf D} = \sum_{k=-1} \varepsilon^{-k}\, {\bf D}_{(-k)}
\ee
where
\be
{\bf D}_{ca}^{(-k)} = 
\begin{dcases}
    ( \vartheta_c | {\cal D} W {\wedge} \varphi_a^{(n)} )_{dW,0} 
    &\text{ if } k=-n,\\
    \textstyle\sum_{l=-n}^k ( \vartheta_c | {\cal D} \varphi_a^{(-l)} )_{dW,k-l} 
    + \sum_{l=-n}^{k+1} ( \vartheta_c | {\cal D} W {\wedge} \varphi_a^{(-l-1)} )_{dW,k-l-1} 
    &\text{ if } -n<k<-1,\\
    \textstyle\sum_{l=-n}^{-1} ( \vartheta_c | {\cal D} \varphi_a^{(1)} )_{dW,-l-1} 
    + \sum_{l=0}^{n} ( \vartheta_c | {\cal D} W {\wedge} \varphi_a^{(0)} )_{dW,l} 
    &\text{ if } k=-1,\\
    \textstyle\sum_{l=0}^{n} ( \vartheta_c | {\cal D}\varphi_a^{(l)} )_{dW,k+l} 
    + \sum_{l=0}^{n} ( \vartheta_c | {\cal D} W {\wedge} \varphi_a^{(l)} )_{dW,k+l+1} 
    &\text{ if } k\geq0.\\
\end{dcases}
\ee
To be concrete let us finish by giving an expression for the connection matrices in terms of the above quantities:
\begin{gather}
{\bf\Omega}_{(1)}^{\intercal} = {\bf C}_{(0)}^{-1} {\bf D}_{(1)},\label{Omega-1}\\
{\bf\Omega}_{(0)}^{\intercal} = {\bf C}_{(0)}^{-1} \left( {\bf D}_{(0)} - {\bf C}_{(-1)} {\bf C}_{(0)}^{-1} {\bf D}_{(1)} \right).\label{Omega-0}
\end{gather}

All computation of Euler characteristics $\chi(M)$ and differential equations below have been double-checked with computational software \texttt{Macaulay2} \cite{M2} and \texttt{FIRE6} \cite{Smirnov:2019qkx} respectively.

\subsection{\label{sec:massless-box}Example I: One-Loop Massless Box}

Let us study arguably the simplest example that illustrates the idea behind this paper in a straightforward manner. We consider a one-loop scalar massless box graph $G_{\text{box}}$, where $P{=}4$ and the set of propagators is given by
\be
\mathsf{D}_1 = \ell^2, \quad
\mathsf{D}_2 = (\ell{+}p_1)^2, \quad
\mathsf{D}_3 = (\ell{+}p_1{+}p_2)^2, \quad \mathsf{D}_4 = (\ell{+}p_1{+}p_2{+}p_3)^2,
\ee
where all external momenta are massless, i.e., $p_i^2 = 0$.
The Symanzik polynomials \eqref{Symanzik-polynomials} are given by
\be
{\cal F} = s x_1 x_3 + t x_2 x_4,\qquad {\cal U} = x_1{+}x_2{+}x_3{+}x_4.
\ee
They depend on the two Mandelstam invariants $s=(p_1{+}p_2)^2$ and $t=(p_2{+}p_3)^2$. We can factor out one of these mass-scales by rescaling $x_a = z_a/(-s)$ and defining $y := t/s$ to be the only kinematic variable. This leaves us with the family of integrals
\be
(-s)^{\varepsilon}\frac{\Gamma({-}2\varepsilon)}{\varepsilon\Gamma(2{-}\varepsilon)} \widehat{I}_{\nu_1,\nu_2,\nu_3,\nu_4} := \int_{\mathbb{R}^4_+} \!\! e^{\varepsilon W}\, \varphi_{\nu_1, \nu_2, \nu_3, \nu_4},
\ee
where we factored out an overall kinematics-independent normalization for later convenience. Setting $\delta_a = \delta$, the potential is given by
\be
W = \log({\cal G}) + \delta \sum_{a=1}^{4} \log z_a \qquad\text{with}\qquad {\cal G} := {-} z_1 z_3 {-} y z_2 z_4 + z_1{+}z_2{+}z_3{+}z_{4}.
\ee
Twisted forms are defined through
\be
\varphi_{\nu_1, \nu_2, \nu_3, \nu_4} := (-s)^{2-|\nu|} \frac{\Gamma({-}2\varepsilon)}{\varepsilon\Gamma(4{-}2\varepsilon{-}|\nu| {-} 4\varepsilon \delta)} \frac{1}{{\cal G}^2} \bigwedge_{a=1}^{4} \frac{z_a^{\nu_a-1} dz_a}{\Gamma(\nu_a {+} \varepsilon\delta)}.
\ee

The moduli space of metrics on the box graph, ${\cal M}_{G_\text{box}} = (\C^\times)^4 - \{ {\cal G}=0 \}$, has the Euler characteristic $|\chi({\cal M}_{G_\text{box}})|=3$, as can be checked with \texttt{Macaulay2} \cite{M2}, and therefore we need to choose three basis forms $\{\varphi_a\}_{a=1}^{3}$ to span $H_{dW}^4$. Following \cite{Henn:2014qga}, we take:
\be
\varphi_1 := y s\,\varphi_{0,1,0,2}, \qquad \varphi_2 := s\, \varphi_{1,0,2,0}, \qquad \varphi_3 := \varepsilon y s^2\, \varphi_{1,1,1,1}.
\ee
The powers of $s$, $y$, and $\varepsilon$ are chosen such that the resulting basis elements depend only on the ratio $y$ and in particular be independent of $s$ and $\varepsilon$ to leading orders in $\delta$. Explicitly, we have
\be
\varphi_1 = \frac{\delta^2 y z_4}{2 z_1 z_3 {\cal G}^2} d^4 z + {\cal O}(\delta^3),\qquad
\varphi_2 = \frac{\delta ^2 z_3}{2 z_2 z_4 {\cal G}^2} d^4 z + {\cal O}(\delta^3),\qquad
\varphi_3 = \frac{y}{{\cal G}^2} d^4z + {\cal O}(\delta).\nn
\ee
For simplicity we also choose the same basis for the dual cohomology $\{\vartheta_a\}_{a=1}^{3} \in H_{-dW}^4$, $\vartheta_a = \varphi_a$, which guarantees that the intersection matrix $\mathbf{C}$, as in \eqref{matrix-C}, starts at order $\varepsilon^0$ in the expansion around $\varepsilon \to \infty$. In this case, equations \eqref{matrix-C} and \eqref{matrix-D} simplify to 
\begin{align}
    {\bf C}_{bc}^{(-k)} &= ( \vartheta_b \vert \vphi_c )_{dW,k},\\
    {\bf D}_{ca}^{(-k)} &=
    \begin{cases}
        (\vartheta_c \vert \mathcal{D}W {\wedge} \vphi_a )_{dW,0}
        & \quad \text{if} \quad k=-1, \\
        (\vartheta_c \vert \mathcal{D}W {\wedge} \vphi_a )_{dW,k+1}
         + (\vartheta_c \vert \mathcal{D} \vphi_a )_{dW,k}
        & \quad \text{if} \quad k\geq0.
    \end{cases}
\end{align}

As the first step in computing intersection numbers we find the critical points given by $dW =0$, whose positions to the order ${\cal O}(\delta^2)$ are given by
\begin{align}
&\Crit(W) = 
\Big\{
    \,\Big(
            1{+}\delta{-}3\delta^2,\;
            {-}\delta{+}(2{+}y)\delta^2,\;
            1{+}\delta{-}3\delta^2,\;
            -\delta {+} (2{+}y) \delta^2 
        \Big),
    \\
    &\qquad\qquad\qquad \,\left(
        -\frac{\delta }{y} {+} \frac{(1{+}2y)\delta^2}{y^2},\; 
        \frac{1{+}\delta {-} 3\delta^2}{y},\; 
        -\frac{\delta }{y} {+} \frac{(1+2y)\delta^2}{y^2},\; 
        \frac{1{+}\delta {-} 3\delta^2}{y} 
    \right),
    \nn\\
    &\medmath{\left( 
        1{+}\frac{(1{+}y)\delta}{y}{-}\frac{(1{+}y)^2\delta^2}{y^2},\;
        \frac{1}{y} {+} \frac{(1{+}y)\delta}{y} {-} \frac{(1{+}y)^2\delta^2}{y^2},\; 1{+}\frac{(1{+}y)\delta}{y}{-}\frac{(1{+}y)^2\delta^2}{y^2},\;
        \frac{1}{y} {+} \frac{(1{+}y)\delta}{y} {-} \frac{(1{+}y)^2\delta^2}{y^2}
    \right)}
\Big\} {+} {\cal O}(\delta^3).\nn
\end{align}
Let us remark that the fact that only the last critical point remains at a non-singular position as $\delta \to 0$ signals that there is only one top-level diagram (e.g. given by $\varphi_3$) in agreement with the program \texttt{Mint} provided in \cite{Lee:2013hzt}.

Computing matrix $\mathbf{C}$ from \eqref{matrix-C} using higher residue pairings gives to leading orders:
\be
\mathbf{C}_{(0)} = \left(
\begin{array}{ccc}
 -\frac{\delta ^2}{4} & 0 & 0 \\
 0 & -\frac{\delta ^2}{4} & 0 \\
 0 & 0 & 1{-}2 \delta ^2 \\
\end{array}
\right) + {\cal O}(\delta^3), \qquad \mathbf{C}_{(-1)} = \left(
\begin{array}{ccc}
0 & 0 & 0 \\
 0 & 0 & 0 \\
 0 & 0 & 0 \\
\end{array}
\right) + {\cal O}(\delta^3).
\ee
Vanishing of the diagonal entries of $\mathbf{C}_{(-1)}$ is actually an exact-$\delta$ statement since we used the same bases for the two cohomologies and the subleading higher residue pairing is anti-symmetric. In order to compute the matrix $\mathbf{D}$ from \eqref{matrix-D} we first need to evaluate how the kinematic space differential ${\cal D} = dy \partial_y$ acts on forms. We have
\be
{\cal D}W \wedge \varphi_a = -\frac{z_2 z_4}{\cal G} dy \wedge \varphi_a,
\ee
as well as
\be
{\cal D}\varphi_a = \left( \frac{\delta_{a,1}{+}\delta_{a,3}}{y} + \frac{2 z_2 z_4}{\cal G} \right) dy \wedge \varphi_a.
\ee
Plugging into the definition \eqref{matrix-D} we find to order ${\cal O}(\delta^2)$:
\be
\mathbf{D}_{(1)} = \left(
\begin{array}{ccc}
 \frac{\delta ^2  }{4 y} & 0 & \frac{\delta ^2}{2 y(y+1)} \\
 0 & 0 & -\frac{\delta ^2 }{2 (y+1)} \\
 \frac{\delta ^2 }{2 y(y+1)} & -\frac{\delta ^2  }{2 (y+1)} & -\frac{  1 + (1+y)(2 - \delta) \delta}{y (y+1)} \\
\end{array}
\right) dy,
\qquad
\mathbf{D}_{(0)} = \left(
\begin{array}{ccc}
 0 & 0 & 0 \\
 0 & 0 & 0 \\
 0 & 0 & 0 \\
\end{array}
\right) dy.
\ee
Finally, we put everything together using \eqref{Omega-1} and \eqref{Omega-0}, which can now be truncated to the finite order in $\delta$ and read
\be
{\bf \Omega}_{(1)} = \left(
\begin{array}{ccc}
	-\frac{1 }{y} & 0 & 0 \\
	0 & 0 & 0 \\
	-\frac{2  }{y(y+1)} & \frac{2 }{y+1} & -\frac{1 }{y(y+1)} \\
\end{array}
\right) dy, \qquad
{\bf \Omega}_{(0)} = \left(
\begin{array}{ccc}
	0 & 0 & 0  \\
	0 & 0 & 0 \\
	0 & 0 & 0 \\
\end{array}
\right) dy.
\ee
Note that even though some entries of the inverse matrix $\mathbf{C}_{(0)}^{-1}$ contain terms of order ${\cal O}(\delta^{-1})$ coming from ${\cal O}(\delta^4)$ of $\mathbf{C}_{(0)}$ not given above, they drop out in the final contraction $\mathbf{C}_{(0)}^{-1} \mathbf{D}_{(1)}$.

Assuming no subsubleading terms contribute, which would violate $\varepsilon$-polynomiality, we can write the differential equations matrix as
\be
{\bf \Omega} = \varepsilon \frac{dy}{y} \left(
\begin{array}{ccc}
	-1 & 0 & 0 \\
	0 & 0 & 0 \\
	-2 & 0 & -1 \\
\end{array}
\right) + \varepsilon \frac{dy}{y{+}1} \left(
\begin{array}{ccc}
	0 & 0 & 0 \\
	0 & 0 & 0 \\
	2 & 2 & 1 \\
\end{array}
\right)
\ee
in agreement with \texttt{FIRE6} and \cite{Henn:2014qga}. The connection has simple poles at $y = 0,-1,\infty$, which physically correspond to singularities at $t{=}0$, $u{=}0$, and $s{=}0$ respectively. This gives a definition of the vector bundle \eqref{section} for the family of box integrals.

\subsection{\label{sec:massive-sunrise}Example II: Two-loop Massive Sunrise}

We consider the two-loop sunrise diagram as our second example. It is known that it integrates to elliptic functions and consequently its differential equation is not homogeneous in $\vep$, see, e.g., \cite{Laporta:2004rb,Remiddi:2013joa,Bloch:2013tra,Adams:2014vja,Kalmykov:2016lxx}. In principle one can rescale basis integrals by periods of elliptic curves to bring the differential equations to an $\vep$-form, see, e.g., \cite{Bogner:2019lfa}. We will not do it here in order to preserve rationality of the connection and illustrate the use of subleading higher residue pairings. In this example, $P{=}3$ and the set of propagators is
\be
\mathsf{D}_1 = \ell_1^2 + m_1^2,
\qquad 
\mathsf{D}_2 =  \ell_2^2 + m_2^2,
\qquad 
{\mathsf D}_3 =  (p {+} \ell_1 {+} \ell_2)^2 + m_3^2
\ee
with non-zero masses, $m_i {\neq} 0$.
The corresponding Symanzik polynomials \eqref{Symanzik-polynomials} are given by
\begin{gather}
    \mathcal{F} = (m_1^2 {+} m_2^2 {+} m_3^2 {+} s) x_1 x_2 x_3  + m_1^2\, x_1^2 \left(x_2{+}x_3\right)+m_2^2\, x_2^2 \left(x_1{+}x_3\right)  +m_3^2\,  x_3^2 \left(x_1{+}x_2\right),\\
    \mathcal{U} = x_1 x_2 + x_1 x_3 + x_2 x_3.
\end{gather}
They depend on four kinematic invariants, $s := p^2$ and $m_1^2, m_2^2, m_3^2$. We can factor out one scale, say $m_1^2$, leaving us with only three kinematic variables $(y_1, y_2, y_3) := (s, m_2^2, m_3^2)/m_1^2$. After rescaling Schwinger parameters using $x_a = z_a / m_1^2$ we obtain the family of integrals
\be\label{sunset-integral}
    m_1^{2\vep}
    \frac{\Gamma(3{-}3\vep)}{\Gamma(2{-}\vep)}
    \widehat{I}_{\nu_1,\nu_2,\nu_3}
    := \int_{\mathbb{R}^3_+} e^{\vep W} 
        \vphi_{\nu_1,\nu_2,\nu_3}
\ee
with the potential given by $W = \log(\mathcal{G}) + \delta \sum_{a=1}^3 \log z_a$, where
\be\label{sunset-G}
\mathcal{G} = \left(1{+}y_1{+}y_2{+}y_3\right) z_1 z_2 z_3 +  z_1^2 \left(z_2{+}z_3\right)+ y_2 z_2^2 \left(z_1{+}z_3\right)+ y_3 z_3^2 \left(z_1{+}z_2\right)+ z_1 z_2 +z_2 z_3 +  z_3 z_1.
\ee
As before, we set $\delta_a {=} \delta$. The corresponding twisted forms read
\be
    \vphi_{\nu_1,\nu_2,\nu_3}
    := (m_1^2)^{|\nu|-3}
    \frac{\Gamma(3{-}3\ep)}{\Gamma(6{-}|\nu|{-}3(1{+}\delta)\vep)}
    \frac{1}{\mathcal{G}^2}
   \bigwedge_{a=1}^3
    \frac{z_a^{\nu_a-1} dz_a}{\Gamma(\nu_a{+}\vep\delta)}.
\ee

The moduli space of metrics on the sunrise graph, $\mathcal{M}_{G_\text{sun}} := (\mathbb{C}^\times)^3-\{\mathcal{G}=0\}$, has Euler characteristic $|\chi (\mathcal{M}_{G_\text{sun}})|=7$ according to \texttt{Macaulay2} \cite{M2}, in agreement with \cite{Kalmykov:2016lxx,Bitoun:2017nre}. Therefore we must choose a basis of seven forms $\{\vphi_a\}_{a=1}^7$ to span $H^3_{dW}$, which we take to be
\begin{gather}\label{basis-1}
    \vphi_1 := (1{-}\vep) \vphi_{1,1,0},
    \qquad
    \vphi_2 := (1{-}\vep) \vphi_{1,0,1},
    \qquad 
    \vphi_3 := (1{-}\vep) \vphi_{0,1,1},
     \\ \label{basis-2}
    \vphi_4 := (1{-}2\vep) \vphi_{1,1,1},
    \qquad
    \vphi_5 := - \vphi_{2,1,1},
    \qquad
    \vphi_6 := - \vphi_{1,2,1},
    \qquad 
    \vphi_7 := - \vphi_{1,1,2},
\end{gather}
where we have chosen normalization constants that will turn out to bring the connection into the form $\boldsymbol\Omega= \boldsymbol\Omega_{(0)} {+} \vep
\boldsymbol \Omega_{(1)}$. Explicitly, to leading orders in $\delta$ we have
\begin{gather}\label{varphi-1}
    \vphi_{a=1,2,3} =  \frac{\delta \vep}{3 z_{4-a} \, {\cal G}^2} d^3z + {\cal O} (\delta^2),
    \\   \label{varphi-2}  
    \vphi_4 =  \frac{1{-}2\vep}{{\cal G}^2} d^3 z + {\cal O} (\delta), \qquad \vphi_{a=5,6,7} = \frac{(2{-}3\vep) z_a}{{\cal G}^2} d^3 z + {\cal O} (\delta).
\end{gather}
For simplicity, we also take 
\begin{align}
    \vartheta_{a=1,2,3} &:= \frac{\vphi_{a}}{\vep},
    \qquad
    \vartheta_4 := \frac{\vphi_4}{1{-}2\vep},
    \qquad
    \vartheta_{a=5,6,7} := \frac{\vphi_{a}}{2{-}3\vep}
\end{align}
as our basis for the dual cohomology $\{\vartheta_a\}_{a=1}^7 \in H^3_{-dW}$, where the normalization is taken to remove the dependence on $\vep$ in the leading $\delta$-order compared to \eqref{varphi-1} and \eqref{varphi-2}, which simplifies power counting. In particular, equations \eqref{matrix-C} and \eqref{matrix-D} become
\begin{align}
    \mathbf{C}_{bc}^{(-k)}
    &= \begin{cases}
        (\vartheta_b \vert \vphi^{(1)}_c )_{dW,0}
        & \quad \text{if} \quad k=-1,
        \\
        (\vartheta_b \vert \vphi^{(1)}_c)_{dW,k+1}
            + (\vartheta_b \vert \vphi^{(0)}_c)_{dW,k}
        & \quad \text{if} \quad k \geq 0,
    \end{cases}
\\
    \mathbf{D}_{ca}^{(-k)}
    &= \begin{cases}
        (\vartheta_c \vert \mathcal{D}W {\wedge} \vphi^{(1)}_a)
        & \quad \text{if} \quad k=-2,
        \\
         (\vartheta_c \vert \mathcal{D}W {\wedge} \vphi^{(1)}_a )_{dW,1}
            + (\vartheta_c \vert (\mathcal{D} \vphi^{(1)}_a 
                + \mathcal{D}W {\wedge} \vphi^{(0)}_a ))_{dW,0}
        & \quad \text{if} \quad k=-1.
    \end{cases}
\end{align}
Note that we have suppressed showing an explicit expression for $\mathbf{D}^{(-k)}_{ca}$ for $k\geq0$ since it is not needed for this example.  

As before, we begin by computing the critical points of the potential $W$. In order to keep the formulae concise we display the result for the equal-mass case, that is $(y_1, y_2, y_3) = (y,1,1)$. We expand it to order $\mathcal{O}(\delta)$, which we found sufficient for the computation of higher residue pairings:
\begin{align}\label{sunset-critical-points}
    \text{Crit}(W) &= 
    \bigg\{
        -\frac{
        \left(
            (y {-} 1)^2 {-} 2 \delta y,\;
            2 (y {-} 1) {+} 2 \delta y,\;
            2 (y {-} 1) {+} 2 \delta y
        \right)
        }{3(y{-}1)(y{+}1)},
 \\ & \qquad 
        -\frac{
        \left(
            2 (y {-} 1) {+} 2 \delta y,\;
            (y {-} 1)^2 {-} 2 \delta y,\;
            2 (y {-} 1) {+} 2 \delta y
        \right)
        }{3(y{-}1)(y{+}1)},
    \nn \\ & \qquad 
        -\frac{
        \left(
            2 (y {-} 1) {+} 2 \delta y,\;
            2 (y {-} 1) {+} 2 \delta y,\;
            (y {-} 1)^2 {-} 2 \delta y
        \right)
        }{3(y{-}1)(y{+}1)},
    \nn \\ & \qquad
        -\frac{2{+}\delta}{y{+}9}
        \left(
            1,1,1
        \right) ,\;
        -\frac{
        \left(
            y {-} 1 {-} \delta,\;
            y {-} 1 {-} \delta,\;
            \delta 
        \right)
        }{3(y{-}1)},
    \nn \\ & \qquad 
    -\frac{
        \left(
            y {-} 1 {-} \delta,\;
            \delta,\;
            y {-} 1 {-} \delta
        \right)
        }{3(y{-}1)},\;
        -\frac{
        \left(
           \delta, \;  
            y {-} 1 {-} \delta,\;
            y {-} 1 {-} \delta
        \right)
        }{3(y{-}1)}
    \bigg\} 
    + \mathcal{O}(\delta^2).\nn
\end{align}
Notice that the coordinates of the first and last three critical points are permutations of each other, which is guaranteed by permutation symmetry of the potential $W$ in the equal-mass case.

At this stage let us comment on counting the size of the basis. In the $\delta \to 0$ limit, only the first $4$ critical points from \eqref{sunset-critical-points} remain at non-singular positions implying that there are $4$ top-level diagrams (e.g. given by $\{\varphi_a\}_{a=4}^{7}$) in agreement with $\texttt{Mint}$ \cite{Lee:2013hzt}. (In the general-mass case the behavior of critical points around $\delta$ remains unchanged and the same conclusion holds.) We also chose our bases of twisted forms in \eqref{basis-1} and \eqref{basis-2} such that $\{\varphi_1, \varphi_2, \varphi_3\}$ as well as $\{\varphi_5,\varphi_6,\varphi_7\}$ are related by relabelling symmetry given by permuting the masses $(m_1, m_2, m_3)$. Including this non-linear relation would mean there are only $3$ independent integrals to compute, and only $2$ in the top-level. In addition, in the equal-mass case the integrals over $\{\varphi_1, \varphi_2, \varphi_3\}$ as well as $\{\varphi_5,\varphi_6,\varphi_7\}$ are equal even without invoking symmetry relations. Let us stress that none of these facts contradicts the result that $\dim H^3_{dW} = 7$. This is because even though the integrals might evaluate to the same function, the corresponding twisted forms are not cohomologous (the reason why such a possibility exists is that the integration contour in \eqref{sunset-integral} is permutation-symmetric and kept constant). The fact that $\{\varphi_a\}_{a=1}^{7}$ provide a basis of twisted cohomology associated to $W$ with \eqref{sunset-G} can be checked by confirming that the matrix $\mathbf{C}$ has full rank.

Following the procedure outlined in Sections \ref{sec:vector-bundles} and \ref{sec:massless-box}, we expand $\mathbf{C}$ and $\mathbf{D}$ to $\mathcal{O}(\vep^0)$. We found it sufficient to keep their expressions up to order ${\cal O}(\delta)$, but given that they are large and not illuminating, we will not spell out the details here. Their contractions give rise to the expansion of $\mathbf{\Omega}$ as in \eqref{Omega-1} and \eqref{Omega-0}. Assuming polynomiality, the result reads in the equal-mass case:
\be
   {\bf \Omega} = \frac{dy}{y} \boldsymbol\omega_0   
        + \frac{dy}{y+1} \boldsymbol\omega_{-1} 
        + \frac{dy}{y+9} \boldsymbol\omega_{-9},
\ee
where
\begin{align}
    \boldsymbol\omega_0 &= 
    \left(
        \begin{array}{ccccccc}
            \SetToWidest{0} & \SetToWidest{0} & \SetToWidest{0} & \SetToWidest{0} & \SetToWidest{0} & \SetToWidest{0} & \SetToWidest{0} \\
            0 & 0 & 0 & 0 & 0 & 0 & 0 \\
            0 & 0 & 0 & 0 & 0 & 0 & 0 \\
            0 & 0 & 0 & 1 & -1 & -1 & -1 \\
            0 & 0 & 0 & 0 & 0 & 0 & 0 \\
            0 & 0 & 0 & 0 & 0 & 0 & 0 \\
            0 & 0 & 0 & 0 & 0 & 0 & 0 \\
        \end{array}
    \right)
    -2 \vep \left(
        \begin{array}{ccccccc}
             \SetToWidest{0} & \SetToWidest{0} & \SetToWidest{0} & \SetToWidest{0} & \SetToWidest{0} & \SetToWidest{0} & \SetToWidest{0} \\
            0 & 0 & 0 & 0 & 0 & 0 & 0 \\
            0 & 0 & 0 & 0 & 0 & 0 & 0 \\
            0 & 0 & 0 & 1 & -1 & -1 & -1 \\
            0 & 0 & 0 & 0 & 0 & 0 & 0 \\
            0 & 0 & 0 & 0 & 0 & 0 & 0 \\
            0 & 0 & 0 & 0 & 0 & 0 & 0 \\
        \end{array}
    \right),
    \\
    \boldsymbol\omega_{-1} &=
    \frac{1}{4}\left(
        \begin{array}{ccccccc}
         \SetToWidest{0} & \SetToWidest{0} & \SetToWidest{0} & \SetToWidest{0} & \SetToWidest{0} & \SetToWidest{0} & \SetToWidest{0} \\
        0 & 0 & 0 & 0 & 0 & 0 & 0 \\
        0 & 0 & 0 & 0 & 0 & 0 & 0 \\
        0 & 0 & 0 & 0 & 0 & 0 & 0 \\
        1 & 1 & -3 & -2 & 4 & 0 & 0 \\
        1 & -3 & 1  & -2 & 0 & 4 & 0 \\
        -3 & 1 & 1 & -2 & 0 & 0 & 4 \\
        \end{array}
    \right)
    - \frac{\vep}{4}
    \left(
        \begin{array}{ccccccc}
         \SetToWidest{0} & \SetToWidest{0} & \SetToWidest{0} & \SetToWidest{0} & \SetToWidest{0} & \SetToWidest{0} & \SetToWidest{0} \\
        0 & 0 & 0 & 0 & 0 & 0 & 0 \\
        0 & 0 & 0 & 0 & 0 & 0 & 0 \\
        0 & 0 & 0 & 0 & 0 & 0 & 0 \\
        1  & 1  & -3 & -3 & 8 & 0 & 0 \\
        1  & -3  & 1  & -3 & 0 & 8 & 0 \\
        -3 & 1  & 1  & -3 & 0 & 0 & 8 \\
        \end{array}
    \right),
    \\
    \boldsymbol\omega_{-9} &=
    \frac{1}{12}\left(
    \begin{array}{ccccccc}
     \SetToWidest{0} & \SetToWidest{0} & \SetToWidest{0} & \SetToWidest{0} & \SetToWidest{0} & \SetToWidest{0} & \SetToWidest{0} \\
    0 & 0 & 0 & 0 & 0 & 0 & 0 \\
    0 & 0 & 0 & 0 & 0 & 0 & 0 \\
    0 & 0 & 0 & 0 & 0 & 0 & 0 \\
    1 & 1 & 1 & -2 & 4 & 4 & 4 \\
    1 & 1 & 1 & -2 & 4 & 4 & 4 \\
    1  & 1  & 1 & -2 & 4 & 4 & 4 \\
    \end{array}
    \right)
    -\frac{\vep}{12}
    \left(
    \begin{array}{ccccccc}
     \SetToWidest{0} & \SetToWidest{0} & \SetToWidest{0} & \SetToWidest{0} & \SetToWidest{0} & \SetToWidest{0} & \SetToWidest{0} \\
    0 & 0 & 0 & 0 & 0 & 0 & 0 \\
    0 & 0 & 0 & 0 & 0 & 0 & 0 \\
    0 & 0 & 0 & 0 & 0 & 0 & 0 \\
    1  & 1  & 1  & -3 & 8 & 8 & 8 \\
    1  & 1  & 1  & -3 & 8 & 8 & 8 \\
    1  & 1  & 1  & -3 & 8 & 8 & 8 \\
    \end{array}
\right),
\end{align}
in agreement with \texttt{FIRE6} \cite{Smirnov:2019qkx}.
The connection has simple poles at $y=0,-1,-9,\infty$ which correspond to singularities at $s=0,-m_1^2,-(3m_1)^2$ and $m_1=0$ respectively.

In the generic-mass case, where $(m_1, m_2, m_3)$ are all distinct, the resulting connection $\mathbf{\Omega}$ becomes more complicated and would not fit within the margins on this paper. Nevertheless, since the kinematic space is $3$-dimensional, we can perform a non-trivial check on integrability of the connection. To be precise, $\mathbf{\Omega}$ takes the form
\be
\mathbf{\Omega} =: \sum_{i=1}^{3} (\widehat{\mathbf{\Omega}}_{(0,i)} + \varepsilon \widehat{\mathbf{\Omega}}_{(1,i)}) dy_i
\ee
and hence the constraint \eqref{flatness} gives more explicitly, for $i,j = 1,2,3$,
\be
\partial_{y_i} \widehat{\mathbf{\Omega}}_{(0,j)}
-\partial_{y_j} \widehat{\mathbf{\Omega}}_{(0,i)}
- [ \widehat{\mathbf{\Omega}}_{(0,i)} , \widehat{\mathbf{\Omega}}_{(0,j)} ] = 0,
\ee
\be
\partial_{y_i} \widehat{\mathbf{\Omega}}_{(1,j)}
-\partial_{y_j} \widehat{\mathbf{\Omega}}_{(1,i)}
- [ \widehat{\mathbf{\Omega}}_{(0,i)} , \widehat{\mathbf{\Omega}}_{(1,j)} ] - [ \widehat{\mathbf{\Omega}}_{(1,i)} , \widehat{\mathbf{\Omega}}_{(0,j)} ] = 0,
\ee
\be
[ \widehat{\mathbf{\Omega}}_{(1,i)} , \widehat{\mathbf{\Omega}}_{(1,j)} ] = 0,
\ee
at orders $\varepsilon^0$, $\varepsilon^1$, and $\varepsilon^2$ respectively. We checked that the result of computing $\mathbf{\Omega}$ with higher residue pairings satisfies these integrability conditions and agrees with $\texttt{FIRE6}$.

\section{\label{sec:discussion}Discussion}

In this work we studied a surprising phenomenon in which the information about Feynman integrals in dimensional regularization around $D=4$ can be fully extracted from a \emph{finite} expansion around saddle points on the moduli space of graphs ${\cal M}_G$. This behavior is in contrast with a more conventional $1/D$ expansion of Feynman integrals, previously studied in the context of gravity \cite{Strominger:1981jg,BjerrumBohr:2003zd,Hamber:2005vc}, which in principle requires an infinite number of corrections to reach $D=4$. We nonetheless hope that a deeper connection between the two approaches can be made in the future.

One of the more intriguing questions is whether there exists an intrinsic property of a basis of twisted forms that could determine if the connection matrix $\mathbf{\Omega}$ in \eqref{Omega-definition} is homogeneous in $\varepsilon$ without direct computations. For instance, it is known that intersection numbers of \emph{logarithmic} forms are always homogeneous in $\varepsilon$ and only the leading higher residue pairing is non-vanishing, see, e.g., \cite{matsumoto1998,Mizera:2019gea}. In the present context we are looking a property of a \emph{basis} of twisted forms, rather than an individual one. It would be fascinating to understand a similar geometric condition that leads to an $\varepsilon$-form differential equations, or decide whether such a basis could even exist. Although we used a representation in terms of Symanzik polynomials, as in \eqref{superpotential}, there is no substantial difficulty in repeating our analysis in other ways, e.g., using the original loop-momentum variables \cite{CHP} or Baikov representation \cite{Mastrolia:2018uzb,Frellesvig:2019uqt,Frellesvig:2019kgj}, where the answer to this question might prove easier.

One of the byproducts of our investigation, which has not been given the attention it deserves, is a new connection between intersection numbers in scattering amplitudes and Landau--Ginzburg models. Indeed, for a given potential $W$ one can define a Hilbert space of such a theory with states given by twisted cohomology classes on $M$. Two-point functions in this model are computed by the intersection numbers $\la \varphi_- | \varphi_+ \ra_{dW}$ and can be expanded in $1/\tau$ using higher residue pairings, see, e.g., \cite{Blok:1991bi}. It would be very interesting to construct concrete realizations of such models in the case of $M{=}{\cal M}_{0,n}$ or $M{=}{\cal M}_G$.

Given that a motivation for the present paper partially came from the scattering equations formalism \cite{Cachazo:2013hca}, let us comment on why we have not studied higher residue pairings on $M{=}{\cal M}_{0,n}$ in more depth. It is known that intersection numbers compute tree-level amplitudes with poles of the type $\frac{1}{p^2 + \Z/\alpha'}$. For concreteness, let us give an example in a simple case of massive cubic scalar theory with $m^2 = 1/\alpha'$, whose $4$-pt amplitude can be written as
\begin{gather}
\Braket{ \frac{d^4z}{\text{vol SL}(2,\C)} | \left(\! \frac{1}{(z_{12} z_{23} z_{34} z_{41})^2} {+} \frac{1}{(z_{13} z_{32} z_{24} z_{41})^2} {+} \frac{1}{(z_{13} z_{34} z_{42} z_{21})^2} \!\right) \!\frac{d^4z}{\text{vol SL}(2,\C)} }_{\!dW} \nn\\
\quad = \frac{1}{s + 1/\alpha'} + \frac{1}{t + 1/\alpha'} + \frac{1}{u + 1/\alpha'}
\end{gather}
in the notation of \cite{Mizera:2019gea}. Clearly, to leading order in the massless limit $\alpha' \to \infty$, the intersection number computes the $4$-pt amplitude of massless scalars, as expected since the leading higher residue pairing coincides with the Cachazo--He--Yuan formula \cite{Cachazo:2013hca}. Alas, the expansion in $1/\alpha'$ does not truncate in the presence of massive propagators and higher residue pairings do not seem terribly useful in this context. Combined with the fact that scattering equations do not have algebraic solutions for $n > 5$, suggests that in order to compute scattering amplitudes on ${\cal M}_{0,n}$ one should instead employ much more efficient recursion relations \cite{Mizera:2019gea} that are exact in $\alpha'$.

Finally, the theory of primitive forms, which gave rise to higher residue pairings, has been developed in order to generalize the classic theory of elliptic integrals to more general spaces \cite{19831231,saito1983higher}. We expect it to play a crucial role in recent developments connecting Feynman integrals to Calabi--Yau geometries \cite{Brown:2009ta,Brown:2010bw,Bourjaily:2018ycu,Vanhove:2018mto,festi2018bhabha,Bourjaily:2018yfy,Besier:2019hqd,Bourjaily:2019hmc}.

\acknowledgments
We thank S.~Caron-Huot, E.~Casali, S.~Li, J.~Maldacena, L.~Mason, P.~Mastrolia, K.~Saito, and P.~Tourkine for useful discussions and correspondence. S.M. thanks the Mathematical Institute, University of Oxford for hospitality during parts of this work. S.M. gratefully acknowledges the funding provided by Carl~P.~Feinberg. 
A.P. thanks the Institute for Advanced Study for its hospitality while this work was completed. A.P. is grateful for funding provided by the Fonds de Recherche du Qu\'ebec.

\appendix
\section{\label{app:Gross-Mende}Gross--Mende Limit and Stokes Phenomena}

In this appendix we briefly clarify the computation of the high-energy ($\alpha' \to \infty$) limit of string theory amplitudes in the simplest example of genus-zero four-point scattering, first studied in \cite{RobertsThesis,Fairlie:1972zz,Gross:1987ar,Gross:1987kza}. We consider fixed-angle scattering, which corresponds to keeping the ratio $s/t$ constant, for massless external kinematics with $s=(p_1{+}p_2)^2$, $t=(p_2 {+} p_3)^2$, $u=(p_1{+}p_3)^2$, and $s{+}t{+}u{=}0$.

We focus on the case of open strings first. Let us consider a contribution coming from Chan--Paton ordering $(1234)$ of vertex operators on the boundary of disk, which in the $\SL(2,\mathbb{R})$-fixing
$(x_1,x_2,x_3,x_4) = (0,x,1,\infty)$ takes the general form
\be\label{Veneziano}
\int_0^1 |x|^{\alpha' s-n}\, |1{-}x|^{\alpha' t-m}\, dx \;=\; \frac{\Gamma(\alpha's{-}n{+}1)\Gamma(\alpha't{-}m{+}1)}{\Gamma(\alpha's{+}\alpha't{-}n{-}m{-}2)},
\ee
where $n,m \in \Z$ are some constants depending of the matter content of vertex operators.

It is a common misconception that the $\alpha' \to \infty$ limit of this amplitude is dominated by a single saddle-point at $x_\ast =s/(s{+}t)$. However, it is easily seen on the physical grounds that this cannot be the case. Consider the kinematic space parametrized by $(s,t) \in \R^2$ and focus on physical string (without tachyons) for the sake of argument. There is an infinite number of resonances in the $s$- and $t$-channels, cf. the explicit expression \eqref{Veneziano}, at $1/\alpha'$ spacing extending to $-\infty$ in both directions. Since there are no poles in the $s,t>0$ quadrant, the asymptotic limit does not have any poles either. However, once we change the direction in which the limit is taken, say from just above the $s$-axis to just below, the asymptotic limit ought to have an infinite number of poles. This signals a Stokes phenomenon, which in fact happens upon changing the sign of either $\Re(s)$, $\Re(t)$, or $\Re(u)$, see, e.g., \cite{Mizera:2019gea}.

The reason for this behavior is that there is, in fact, an \emph{infinite} number of saddles that can contribute to the asymptotics of \eqref{Veneziano}. Depending on the direction in the $(s,t)$-plane in which the limit is taken (or alternatively position of $x_\ast$ with respect to the integration contour), only a subset of them might dominate. All such saddles contribute with the same magnitude but distinct phases. As we shall see, this allows for an infinite number of saddles-point contributions to be resummed into a simple oscillatory term.

In order to analyze the asymptotic behavior of \eqref{Veneziano} we first analytically continue its integrand to a complex variable $z$,
\be\label{A2}
\int_{0}^{1} z^{\alpha' s - n}\, (1{-}z)^{\alpha' t -m}\, dz.
\ee
The key observation is that the integrand of \eqref{A2} is no longer defined on the moduli space $\M_{0,4} = \{ z \in \CP^1 \,|\, z \neq 0,1,\infty \}$, but rather on its universal cover $\widetilde{\M}_{0,4}$.\footnote{The space described by the number of windings $(p,q) \in \Z^2$ is in fact the maximal Abelian cover of ${\cal M}_{0,4}$, which is smaller than the universal cover $\widetilde{{\cal M}}_{0,4}$ (covering group of the former is the $1$-st homology group $H_1({\cal M}_{0,4},\Z)$ as opposed to the fundamental group $\pi_1({\cal M}_{0,4})$, cf. \cite{Witten:2013pra}), but is sufficient for our purposes.} This is the case because the part of the integrand $z^{\alpha' s} (1{-}z)^{\alpha' t}$ is multi-valued and hence defines a infinitely-sheeted surface $\widetilde{\M}_{0,4}$. For example, going around the branching point $z{=}0$ in an small anti-clockwise circle $p$ times the integrand changes to $e^{2\pi i \alpha' p s}$ times its original value, and likewise going around $z{=}1$ $q$ times multiplies it by $e^{2\pi i \alpha' q t}$. As a result we find infinite number of saddle points characterized by the number of windings $(p,q)$ around $z{=}0$ and $z{=}1$ respectively (note that winding $r$ times around $z{=}\infty$ is equivalent to $(p,q)=(-r,-r)$ and thus not independent).

There are several ways of consistently computing contributions from all the saddles. For example, we can consider a change of variables 
\be
z = e^u,\qquad 1{-}z = e^v, \qquad\text{such that}\qquad e^u + e^v = 1.
\ee
In order to impose the last constrain we introduce a Lagrange multiplier $w$, so that \eqref{A2} becomes
\be\label{A4}
\alpha' \int_{\Gamma} e^{(\alpha' s - n + 1) u + (\alpha' t - m + 1) v + 2\pi i \alpha' w (e^u + e^v - 1)}\, du\, dv\, dw,
\ee
where $\Gamma = \mathbb{R}_-^2 \times \mathbb{R}$. Solving the critical point equations
\be\label{change-of-variables}
s + 2\pi i w e^u = 0, \qquad t + 2\pi i w e^v = 0, \qquad 2\pi i (e^u + e^v - 1) = 0,
\ee
yields an infinite number of solutions given by
\be
u_{\ast} = \log\left(\frac{s}{s{+}t}\right) + 2\pi i p, \qquad v_{\ast} = \log\left(\frac{t}{s{+}t}\right) + 2\pi i q, \qquad w_{\ast} = \frac{i}{2\pi} (s{+}t),
\ee
for every $(p,q) \in \Z^2$ which are the winding numbers introduced above. Indeed, it is easily seen that undoing the change of variables \eqref{change-of-variables} each $(u_\ast,v_\ast,w_\ast)$ is mapped to the same $z_\ast = s/(s{+}t)$, which is why it ``looked'' like a single saddle point to begin with.

In order to discern which saddles contribute to the $\alpha' \to \infty$ limit, one should first find all the cycles of steepest descent (Lefschetz thimbles) ${\cal J}_{p,q}$ and ascent ${\cal K}_{p,q}$ of the Morse function
\be
\Re(W) = \Re( s u + t v + 2\pi i w (e^u + e^v - 1) )
\ee
emanating from each saddle-point labelled by $(p,q)$. Clearly, if the integral \eqref{A4} was over such ${\cal J}_{p,q}$, it would have been dominated by the $(p,q)$-th saddle in the $\alpha' \to \infty$ limit. Thus, it remains to translate the original integration over $\Gamma$ into those over the steepest descent paths, which can be done by using the relation
\be\label{R3-decomposition}
\eqref{A4} = \alpha'\!\!\! \sum_{(p,q)\in\Z^2} \la \Gamma | {\cal K}_{p,q} \ra \int_{{\cal J}_{p,q}} \!\!\! e^{(\alpha' s - n + 1) u + (\alpha' t - m + 1) v + 2\pi i \alpha' w (e^u + e^v - 1)}\, du\, dv\, dw.
\ee
Here $\la \Gamma | {\cal K}_{p,q} \ra \in \Z$ is the homology intersection number of the corresponding cycles, see \cite{pham1983vanishing,arnold2012singularities} for standard references. Using saddle-point approximation and separating the $(p,q)$-dependent terms, in the high-energy limit we find
\be\label{A9}
\lim_{\alpha' \to \infty} \eqref{A4} = \underbrace{\left( \sum_{(p,q)\in\Z^2} \la \Gamma | {\cal K}_{p,q} \ra\, e^{2\pi i \alpha' (p s + q t)} \right)}_{=: f(s,t)} \frac{e^{(\alpha' s - n + 1)\log\left(\frac{s}{s+t}\right) + (\alpha' t - m +1)\log\left(\frac{t}{s+t}\right)}}{\sqrt{\frac{\alpha' s t}{2\pi (s+t)}}}.
\ee
Crucially, the shape of each ${\cal J}_{p,q}$ and ${\cal K}_{p,q}$ might change drastically depending on values of the parameters $s$ and $t$. For instance, when $s,t>0$, the original integration domain $\Gamma$ is already homologous to ${\cal J}_{0,0}$ and the sum $f(s,t)$ in \eqref{R3-decomposition} has only one term equal to $1$. As a consequence, only a single saddle contributes to the high-energy limit. Note that it is exactly the one that in the original variables lies on the integration contour, i.e., $z_\ast \in (0,1)$.

Nevertheless, in a generic kinematic region, none of the steepest descent cycles equals to $\Gamma$ and the sum in \eqref{R3-decomposition} generically involves an infinite number of terms and consequently an infinite number of saddles contribute to the high-energy limit. As we will see, in those cases $f(s,t)$ can be resumed into a concise expression.\footnote{Finding ${\cal K}_{p,q}$ and their respective intersection numbers $\la \Gamma | {\cal K}_{p,q}\ra$ is a generalization of a similar problem considered in \cite{10.2307/51844,doi:10.1098/rspa.1994.0158} for the gamma function $\Gamma(s)$. As a matter of fact, we could have used the result
\be
\lim_{\alpha' \to \infty}\Gamma(\alpha' s) = \sqrt{\frac{2\pi}{\alpha' s}} e^{\alpha' s(\log(\alpha' s)-1)}\times \begin{dcases}
1 &\qquad s>0,\\
\frac{1}{e^{2\pi i \alpha' s}-1} & \qquad s<0,
\end{dcases}
\ee
already on the Veneziano amplitude on the right-hand side of \eqref{Veneziano}, though this approach would not give us any intuition about generalizations to higher-point or higher-genus amplitudes that cannot be expressed in terms of gamma functions.
}

There is however a different approach to this problem, using homologies with local coefficients (or twisted homologies, as they were called in the main text), which computes $f(s,t)$ in one go. In a nutshell, it allows to ``collapse'' the information about all the branches of $\widetilde{\cal M}_{0,4}$ by endowing each integration cycle with a coefficient of the form $e^{2\pi i (ps+qt)}$ for a given $p,q$. This allows for computations directly on ${\cal M}_{0,4}$, which are much easier than those on the covering space. In this language $f(s,t)$ becomes a single \emph{twisted} intersection number, which can be easily computed \cite{Mizera:2019gea}. In fact, following a computation from Appendix A of \cite{Mizera:2019gea} we find in the physical region $s{<}0,\; t,u{>}0$
\be\label{fst}
f(s,t) = \frac{e^{-2\pi i \alpha' t} - e^{2\pi i \alpha' s}}{1 - e^{2\pi i \alpha' s}} = e^{-2\pi i \alpha' t}\sum_{p=0}^{\infty} e^{2\pi i \alpha' p s} - \sum_{p=1}^{\infty} e^{2\pi i \alpha' p s},
\ee
which in the second equality we rewrote as a sum over the lattice $(p,q) \in \Z^2$, as in \eqref{A9}, from which one can read-off integer coefficients of each $e^{2\pi i \alpha' (ps+qt)}$. Performing similar computations in other kinematic regions it is easily confirmed that in all of them, except for $s,t{>}0$, an infinite number of saddles contribute. Note that $f(s,t)$ is in general not real, but works out so that the whole right-hand side of \eqref{A9} remains real (i.e., compensates for the fact that $\log(-x) = \log(x) + i\pi$ for $x{>}0$ in the exponential). One can check that in the region $t{<}0$, $s,u{>}0$ the factor $f(s,t)$ is obtained from \eqref{fst} by exchanging $s \leftrightarrow t$, which is a consequence of crossing-symmetry of the Veneziano amplitude.

To summarize, in the physical region we have
\begin{align}\label{result}
\lim_{\alpha' \to \infty} \int_0^1 |x|^{\alpha' s-n}\, &|1{-}x|^{\alpha' t-m}\, dx \;=\; (-1)^m \frac{\sin(\pi \alpha' u)}{\sin(\pi \alpha' s)} \sqrt{\frac{-2\pi u}{\alpha' s t}}\\
&\times \exp \left((\alpha's{-}n{+}1) \log (-s) + (\alpha't{-}m{+}1) \log (t) + (\alpha'u{+}n{+}m{-}2) \log (u)) \right).\nn
\end{align}
As remarked before, there are many other indirect ways of obtaining the same result. For instance, one can notice that the no matter which ray in the kinematic space we are considering, the critical point $x_\ast$ is always on the integration contour of \emph{some} partial amplitude. For instance, in the physical region we have $1< x_\ast < \infty$. We can then use two independent monodromy relations \cite{Plahte:1970wy} to solve for the integral over $(0,1)$ in terms of that over $(1,\infty)$. In doing so one recovers the result \eqref{result}, where the oscillatory phases come from solving the monodromy relations. See \cite{Lee:2015wwa} for related discussion. This approach, however, does not seem to scale well to higher-point amplitudes, as Lefschetz thimbles generically do not coincide with open-string integration cycles.

The discussion of closed-string scattering is almost identical, except for an additional step at the beginning. One starts by \emph{homologically splitting} the corresponding complex integral into two copies of integrals over Lefschetz thimbles. In doing so one encounters a homological intersection number of a similar oscillatory type as that in \eqref{fst}. Since Lefschetz thimbles themselves depend on the direction in which the $\alpha' \to \infty$ is taken, so does the asymptotic behavior of amplitudes, which is generically dominated by an infinite number of saddles. We refer the reader to Appendix A of \cite{Mizera:2019gea} for details of this computation.

\bibliographystyle{JHEP}
\bibliography{references}

\end{document}